\documentclass[journal]{IEEEtran}

\usepackage[T1]{fontenc}
\usepackage[cmintegrals]{newtxmath}
\usepackage{bm}
\usepackage{amsmath}
\usepackage[pdftex]{graphicx}
\usepackage{epstopdf}
\usepackage{graphics}
\usepackage{color}
\usepackage{setspace}
\usepackage{enumerate}
\usepackage{multicol,lipsum}
\usepackage{fixltx2e}

\newtheorem{theorem}{Theorem}[section]
\newtheorem{problem}[theorem]{Problem}
\newtheorem{remark}[theorem]{Remark}
\newtheorem{assumption}[theorem]{Assumption}
\newtheorem{proposition}[theorem]{Proposition}
\newtheorem{definition}[theorem]{Definition}
\newtheorem{lemma}[theorem]{Lemma}

\ifCLASSINFOpdf
\else
\fi
\begin{document}
%
\title{Multi-Level Power-Imbalance Allocation Control for Secondary Frequency Control of Power Systems}
%
%
%

\author{Kaihua Xi,
        Hai Xiang Lin,
        Chen Shen,~\IEEEmembership{Senior~member,~IEEE,}
        Jan H. van Schuppen,~\IEEEmembership{Life~member,~IEEE}
      
\thanks{Kaihua Xi, Hai Xiang Lin and Jan H. van Schuppen are with Delft Institute 
of Applied Mathematics, Delft University of Technology, 2628CD, Delft, The Netherlands (e-mail: K.Xi@tudelft.nl).}
\thanks{Chen Shen is with the State Key Laboratory of Power Systems, Department of Electrical 
Engineering, Tsinghua University, Beijing, 10084, China.}
}

\maketitle

\begin{abstract}
A consensus-control-based multi-level control law named \emph{Multi-Level Power-Imbalance Allocation Control} (MLPIAC) is presented for a large-scale power 
system partitioned into two or more areas. 
Centralized control is implemented in each area while distributed control is 
implemented at the coordination level of the areas. Besides
restoring nominal frequency with a minimal control cost,
MLPIAC can improve the transient performance 
of the system through an accelerated convergence of the control inputs without
oscillations. At the coordination level of the control areas, 
because the number of the areas is smaller 
than that of nodes, MLPIAC is more effective to obtain the minimized 
control cost than the purely distributed control law. 
At the level of the control in each area, because the number of nodes 
is much smaller than the total number of nodes in the whole network,
the overheads in the communications and the computations are reduced compared to the 
pure centralized control. The asymptotic 
stability of MLPIAC is proven using the Lyapunov method and 
the performance is evaluated through simulations. 
\end{abstract}


%
\IEEEpeerreviewmaketitle


\section{Introduction}\label{introduction}
Power systems need to be controlled to provide alternating current with nominal frequency,
60 Hz in the USA, and 50 Hz in the main parts of Europe and in Asia.
The power demand fluctuates continuously
due to switching on or off loads.
Consequently, the frequency of the power systems also fluctuates continuously.
So the power grids must be controlled to maintain the
frequency as close as possible to the agreed reference value.
\par
There are three forms of control:
primary, secondary, and tertiary frequency control which are divided based on fast to slow time-scales. Primary 
frequency control synchronizes the frequencies of synchronous machines and balances
the power supply and demand of the power system at a small time-scale. However,
the synchronized frequency may deviate from its nominal 
value. Secondary frequency control restores the nominal frequency at 
a medium time-scale. With a prediction of power demand, tertiary control calculates the 
operating point stabilized by primary and secondary control at a large time-scale, which concerns 
the security and economy of the power system.

The focus of this paper is on secondary frequency control.
We consider power systems with lossless transmission lines,
which comprise traditional synchronous machines, frequency dependent 
devices, e.g., power inverters of renewable energy or frequency dependent 
loads, and passive loads. The control objectives can be 
described as: restore the frequency to its nominal value and minimize the 
control cost of the power system. 
The initial approach to control synthesis of secondary frequency control
is to apply centralized control \cite{Dorfler2016,TRIP201710102,PIAC3}, where a central controller collects state information via communication network and computes the control inputs for the local
actuators. The minimized 
control cost is achieved by solving an economic power dispatch problem. 
In practice today's power systems are becoming so large that they cannot be effectively controlled
by a centralized controller.
The communication overhead and the control computations carried out at the central controller take so much time
that the control objectives cannot be satisfactorily achieved. 
In addition, the centralized control may also not satisfy the requirement 
of the system integrated with a large amount of distributed power sources. 
\par
Therefore, a form of distributed control
is proposed for control of power systems,
 which are either based on passivity method \cite{Survey_Distributed,XiangyuWu2016,Liu2016,Trip2016240,Simpson-Porco2015,Schiffer2017,Shafiee2014} or primal-dual method \cite{Zhao2015,EAGC,UC,Wang2017}. 
In distributed control of a power system, a number of controllers
try to achieve the control objectives of the entire network via coordination and cooperation.
The state information and control inputs are collected and computed by the local controller at each node. 
In order to minimize the  control cost, the controllers 
need to exchange control information with their neighboring controllers via the 
communication network. However, this suffers from a slow convergence to the optimal 
steady state because of the large scale of the power system. 

In addition, these centralized and distributed control usually 
focus on the steady state only. The transient performance is seldom considered when designing 
the control algorithms. During the transient phase, 
\emph{extra frequency oscillations} or \emph{slow convergence} to the steady state
may be introduced due to the control algorithms \cite{Recent_survey_2005,overshoot_report1,overshoot_report2,PIAC3},
which should be avoided for a high quality power supply by the power systems with various 
disturbances. The traditional way to improve the transient performance is
to tune the control gain coefficients through eigenvalue or $\mathcal{H}_2$/$\mathcal{H}_{\infty}$ norm analysis \cite{XiangyuWu2017,Hassan_Bevrani,Hinfinty_control_overshoot}. 
However, besides the complicated computations, these methods 
focus on the linearized system only 
and the improvement of the transient performance is still limited because it also depends on the structure of the control algorithms. In order to improve 
the transient performance, sliding mode based control laws, e.g., \cite{overshoot2,overshoot3},
and fuzzy control based control laws, e.g., \cite{overshoot1}, are proposed, 
which are able 
to shorten the transient phase without the extra oscillations. However,
those control laws use either centralized control or decentralized control without
considering economic 
power dispatch.

The power system usually has a multi-level structure \cite{Kundur1994}.
For example, the systems at the level of communities 
are subsystems of the systems at the level of provinces, which 
are further subsystems at a higher level of the states.  
Control methods for this kind of multi-level power systems are seldom 
considered.  

In this paper, we aim to synthesize a multi-level control law for secondary 
frequency control of large-scale power systems with a multi-level structure, 
which is able to balance the 
advantages and disadvantages of the centralized and distributed control,
and eliminates the extra frequency oscillations and the slow convergence problem. 
The control objectives can be described as: restore the frequency to its nominal value,
prevent the extra frequency oscillations caused by the controllers,
and minimize the economic cost in the operation of the power system.

The contributions of this paper include: 
\begin{enumerate}[(i)]
\item a multi-level control law, \emph{Multi-Level Power-Imbalance Allocation Control}
(MLPIAC), is proposed, which is not only able to balance the advantages and disadvantages of centralized and distributed control, but also suitable for power systems with 
a multi-level structure. 
\item  the control cost is minimized and both the transient performance of  
the frequency and of the control cost can be improved by tuning three gain parameters in MLPIAC.
\item the Lyapunov stability analysis and case study are provided to evaluate 
the asymptotic stability and performance of MLPIAC. 
\end{enumerate}
Not discussed because of limitations of space are robustness of the closed-loop system,
the interaction between frequency control and voltage control,
and the time delays in the communications and measurement of the frequency.  

This paper is organized as follows. We describe
the dynamic model of the power network and formulate 
the problem in Section \ref{Sec:model}. We synthesize MLPIAC with transient performance analysis in Section 
\ref{Sec:Multi}. The asymptotic stability is analysed in Section \ref{Sec:stability} and 
the performance of MLPIAC is evaluated in the case study 
in Section \ref{Sec:case}. Concluding remarks are given in Section \ref{Sec:conclustion}.

\section{Dynamic model and secondary control}\label{Sec:model}

A power network is described by a graph $\mathcal{G}=(\mathcal{V},\mathcal{E})$ with nodes $\mathcal{V}$ and edges
$\mathcal{E}\subseteq \mathcal{V}\times \mathcal{V}$ where a node represents a bus and edge $(i,j)$ represents
 the direct transmission line connection between node $i$ and $j$. The buses may connect to 
 synchronous machines, frequency dependent power sources (e.g., power inverters of renewable energy) or loads, or passive loads. 
 The resistance of the transmission lines are neglected and the susceptance is denoted by 
 $\hat{B}_{ij}$. Denote the set of the buses of the synchronous machines, frequency dependent power sources, passive 
 loads by $\mathcal{V}_M,\mathcal{V}_F,\mathcal{V}_P$ respectively, thus $\mathcal{V}=\mathcal{V}_M\cup\mathcal{V}_F\cup\mathcal{V}_P$.

 The dynamic model of the power system is described by 
 the following \emph{Differential Algebraic Equations} (DAEs), e.g.,\cite{Dorfler2016}, 
\begin{subequations}\label{eq:system1}
\begin{align}
  \hspace{-6pt}
  \dot{\theta}_i&=\omega_i,~i\in\mathcal{V}_M\cup\mathcal{V}_F,\\
M_{i}\dot{\omega}_{i}&=P_{i}-D_{i}{\omega}_{i}-\sum_{j\in \mathcal V}{B_{ij}\sin{(\theta_{i}-\theta_{j})}}+u_{i}, 
i\in \mathcal{V}_M,\label{eq:syn}\\
0&=P_{i}-D_{i}\omega_{i}-\sum_{j\in \mathcal V}{B_{ij}\sin{(\theta_{i}-\theta_{j})}}+u_{i}, i\in \mathcal{V}_F,\label{eq:frq}\\
0&=P_{i}-\sum_{j\in \mathcal V}{B_{ij}\sin{(\theta_{i}-\theta_{j})}}, i\in \mathcal{V}_P, \label{eq:pass} 
 \end{align}
\end{subequations}
where $\theta_i$ is the phase angle at node $i$, $\omega_i$ is the frequency deviation from the nominal 
frequency, i.e., $f^*=50$ or 60 Hz, $M_i>0$ is the moment of inertia of the synchronous machine, 
 $D_i$ is the droop control coefficient, $P_i$ is the 
power generation (or demand), $B_{ij}=\hat{B}_{ij}V_iV_j$ is the effective 
susceptance matrix, $V_i$ is the voltage, $u_i$ is the secondary control input. 
We assume that the nodes participating in secondary frequency control are equipped with the primary controller. 
 Denote the set of the nodes equipped with the secondary controllers by
 $\mathcal{V}_{K}$, thus $\mathcal{V}_{K}= \mathcal{V}_{M}\cup \mathcal{V}_{F}$. 
Since the control of the voltage and the frequency can be decoupled, we do not model the dynamics
of the voltages and assume 
the voltage of each bus is a constant which can be derived from power flow calculation \cite{Kundur1994}. In practice, the voltage can be well controlled 
by Automatic Voltage Regulator (AVR). 
This model and the ones with linearized sine function are
widely studied, e.g.,\cite{Dorfler2016,Dorfler2014,Trip2016240,DePersis2016,Schiffer2017,UC,EAGC},
in which the frequency dependent nodes are usually used to model 
the renewable power inverters. 


The system (\ref{eq:system1}) synchronizes at an equilibrium state,  called \emph{synchronous state} defined as follows \cite{Dorfler20141539}. 
\begin{definition}
A steady state of the power system (\ref{eq:system1}) with constant power loads (generation) yields, 
\begin{subequations}\label{eq:synchronize}
\begin{align}
\omega_i&=\omega_{syn},~i\in\mathcal{V}_M\cup\mathcal{V}_F\\
\dot{\omega}_i&=0,~i\in\mathcal{V}_M\cup\mathcal{V}_F,\\
\theta_i&=\omega_{syn} t+\theta^*_i,~i\in\mathcal{V},\\
\dot{\theta}_i&=\omega_{syn},
\end{align}
\end{subequations}
where $\theta^*_i$ is the phase angle of node $i$ at the steady state, 
$\omega_{syn}\in\mathbb{R}$ is the synchronized frequency deviation. 
\end{definition}

Note that the angle differences $\{\theta^*_{ij}=\theta^*_i-\theta^*_j,~i,j\in\mathcal{E}\}$ determine the power flows in the 
transmission lines. 
Substituting (\ref{eq:synchronize})
into the system (\ref{eq:system1}), we derive the explicit formula of the synchronized frequency deviation as
\begin{eqnarray}
 \omega_{syn}=\frac{\sum_{i\in \mathcal{V}}{P_i}+\sum_{i\in\mathcal{V}_K}{u_i}}{\sum_{i\in \mathcal{V}_M \cup\mathcal{V}_F }{D_i}}. \label{eq:sync}
\end{eqnarray}
The \emph{power imbalance} is defined as $P_s=\sum_{i\in \mathcal{V}}{P_i}$. 
In order to avoid damages to electrical devices in the system, the frequency deviation should be zero, i.e., $\omega_{syn}=0$,
for which the necessary condition is 
$$P_s+\sum_{i\in\mathcal{V}_K}{u_i}=0,$$
which can be satisfied by the set $\{u_i,~i\in\mathcal{V}_K\}$ of control inputs determined by a control law. 
We aim to synthesize an effective secondary frequency control law to control $\omega_{syn}$ to zero with a minimal control cost, that requires solving 
the following economic power dispatch problem,
\begin{eqnarray}
 &&\min_{\{u_i\in \mathbb{R},i\in\mathcal{V}_K\}} \sum_{i\in\mathcal{V}_K }\frac{1}{2}\alpha_iu_i^2,
 \label{eq:optimal1}\\
 \nonumber
 &&s.t. ~~P_s+\sum_{i\in\mathcal{V}_K }{u_i}=0,
\end{eqnarray}
where $\alpha_i\in\mathbb{R}$ is a positive constant and denotes the control price of node $i$ \cite{Wood}.  

Regarding to the existence of a feasible solution of the optimization problem (\ref{eq:optimal1}), 
we make the following assumption.
\begin{assumption}\label{assumption1}
Consider the system (\ref{eq:system1}) with the economic power dispatch problem (\ref{eq:optimal1}), assume 
\begin{enumerate}[(i)]
\item the power supply and demand are constant in a small time interval, thus $P_s$ is a constant. 
\item the power imbalance can be compensated by the control inputs such that
$$-P_s\in\Big[\sum_{i\in\mathcal{V}_K}\underline{u}_i,\sum_{i\in\mathcal{V}_K}\overline{u}_i\Big]$$
\end{enumerate}
\end{assumption}
In practice, the power demand is not constant but continuously fluctuating 
due to the uncertain behaviour of consumers \cite{PIAC2}. However, for the synthesis of the control laws, the power demand is usually assumed as a constant \cite{Ilic2000} as in Assumption \ref{assumption1} (i). On the power supply side, 
tertiary control calculates the operating point of $P_i$ in a small time interval, which is 
stabilized by secondary frequency control. In general, the electricity 
demand can be satisfied. So Assumption \ref{assumption1} (ii) is realistic. 

A necessary condition for the optimal solution of (\ref{eq:optimal1}) is 
\begin{eqnarray}\label{eq:kkt}
\alpha_i u_i=\alpha_j u_j=\lambda,~i,j\in\mathcal{V}_K
\end{eqnarray} 
where $\alpha_iu_i$ is the \emph{marginal cost} of node $i$ and 
$\lambda$ is the \emph{nodal price}. 
With this condition, if the power-imbalance $P_s$ is known, the optimization problem (\ref{eq:optimal1})
can be solved analytically. However, $P_s$ is unknown in practice since the loads 
cannot be predicted precisely in a short time interval. 

After a disturbance, the state of the power system experiences two phases: a transient phase and 
a steady phase. In a centralized 
control law, e.g., \cite{Dorfler2016,PIAC3}, the nodal price estimated by the central controller is 
broadcast to the local controllers which calculate the control inputs according to (\ref{eq:kkt}). So the marginal costs are all identical during the transient phase. However, the central controller 
suffers from the communication overhead with the local controllers and intensive computations. 
In distributed secondary frequency control, the principle of 
consensus control is used to let the marginal costs of all nodes achieve a consensus 
at the steady state \cite{Andreasson2013}. Because the nodes communicate with their neighbours and compute 
the control input locally, the communications and computations are greatly reduced. However, 
this sacrifices the performance of the marginal costs, i.e., they are not identical during 
the transient phase leading to increase in the control cost. Furthermore, the consensus speed decreases as the scale of the network increases,
which further increases the control cost.  
%

The existing control laws usually consider the 
steady state only.
Extra frequency oscillations or slow convergence to the steady 
state may be introduced by the controllers due to the control algorithms
\cite{Recent_survey_2005,overshoot_report1,overshoot_report2}, which deteriorate the transient performance of 
the system. 
%
The dynamics of actuators of the power system may be included into 
the model (\ref{eq:system1}) \cite{Trip2016}. The extra frequency 
oscillation may be reduced by having 
a slow actuator dynamics but then the frequency deviation 
will be large from the nominal value. If the actuator 
dynamics is very fast then there is little difference from 
the case without actuator dynamics. Therefore,
the actuator dynamics does not eliminate the extra oscillation and 
large frequency deviation which are mainly due to the control algorithms.  

In this paper, besides the optimization problem (\ref{eq:optimal1}) for secondary frequency control,
we also focus on the transient performance of the power system.  We consider 
the following problem. 
\begin{problem}\label{problem:paper}
Consider a large-scale power system described by (\ref{eq:system1}). Design a secondary 
frequency control law so as to improve the transient performance with a minimal control cost, i.e., 
eliminate the extra frequency oscillation caused by the oscillations of 
the control inputs and 
accelerate the convergence of the optimization of control cost.    
\end{problem}

In order to eliminate the extra frequency oscillation, a centralized 
control law, \emph{Power-Imbalance Allocation Control} (PIAC) has 
been proposed in \cite{PIAC3}.
In the next Section, we propose a multi-level control law which is a trade-off between centralized and distributed control and thus 
solve Problem \ref{problem:paper}. 
In order to solve the optimization problem (\ref{eq:optimal1}), a communication network connecting 
the controllers is required, for which we make the following assumption.
\begin{assumption}\label{assumption_comm}
Consider the power system (\ref{eq:system1}), assume all the nodes
in the set $\mathcal{V}_K$ are connected by a communication network. 
\end{assumption}

\section{Multilevel control approach}\label{Sec:Multi}

First, with respect to the transient performance of frequency, we decompose 
the frequency deviation into two types, i.e., global frequency deviation and 
relative frequency deviation between the nodes. The transient performance of relative frequency deviation can be 
improved by primary control \cite{Motter2013}. So we focus on improving the transient performance 
of the global frequency deviation. It can be observed 
from (\ref{eq:sync}) that the global frequency deviation at the steady state is determined by 
the power-imbalance and the total damping of the system. Because of the 
heterogeneity of $M_i$ and $D_i$, it is hard to qualify the performance 
of the global frequency deviation during the transient phase. To treat the transient performance 
we define an abstract frequency deviation to measure the global frequency deviation as follows.
\begin{definition}
 For the power system (\ref{eq:system1}), define an abstract frequency deviation $\omega_s$ \cite{PIAC3} as follows, 
 \begin{eqnarray}\label{eq:global}
 M_s\dot{\omega}_s=P_s-D_s\omega_s+u_s
\end{eqnarray}
where $M_s=\sum_{i\in\mathcal{V}_M}{M_i}$ denotes effective inertia, $D_s=\sum_{i\in\mathcal{V}}{D_i}$ denotes
effective droop control coefficient, and $u_s=\sum_{i\in\mathcal{V}_K}u_i$ denotes the total control input of the system. 
\end{definition}

$\omega_s(t)$ is different from $\omega_i(t)$ and $\omega_{syn}(t)$.
It can be easily obtained that $\omega_s^*=\omega_{syn}=\omega_i$ at the steady state. 
 The dynamics of $\omega_s$ involves
the total inertias, droop control coefficients and power-imbalance, so it is reasonable 
to use $\omega_s$ to study the transient performance of the global frequency deviation.

Second, regarding to the distributed control, we partition the network denoted by $A$ into 
$m$ areas such that 
\begin{subequations}\label{partition}
 \begin{align}
  &A=A_1\cup\cdots \cup A_r\cup\cdots \cup A_m, ~Z_m=\{1,2,\cdots,m\},\\
  &A_r\cap A_q=\emptyset, ~r\neq q,~r,~q\in Z_m. 
 \end{align}
\end{subequations}
There is an area controller in each area. 
Denote the set of the nodes in area $A_r$ by $\mathcal{V}_{A_r}$, the nodes of 
the synchronous machines by $\mathcal{V}_{M_r}$, the nodes of the frequency dependent power sources
by $\mathcal{V}_{F_r}$, the nodes with secondary controller by $\mathcal{V}_{K_r}$, the marginal cost of 
area $A_r$ by $\lambda_r$, and the set of the neighboring areas of $A_r$ connected by communication line
as $Z_{m_r}$. 

\begin{figure}[ht]
 \includegraphics[scale=0.7]{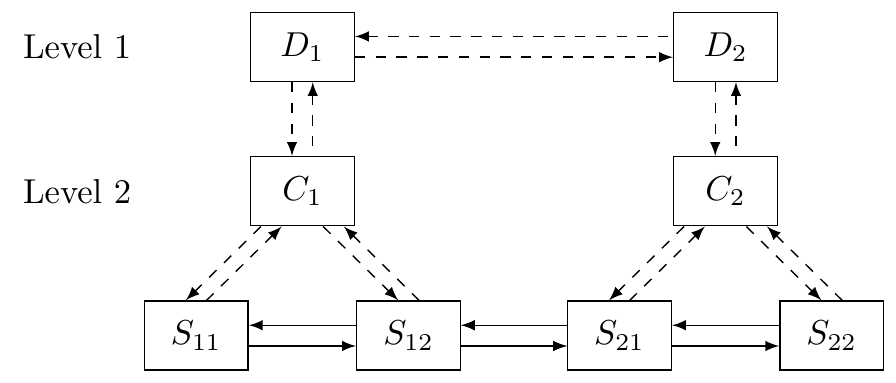}
 \caption{Diagram of the multi-level control of power systems. The dashed lines denote
 the communications and the solid lines denote the physical coupling between the subsystems. }
 \label{hybrid}
\end{figure}

We refer to level 1 for the coordination control of all areas, level 2 for the 
control of an area, and level 3 for the primary control of a node. 
 MLPIAC focuses on the secondary frequency control at level 1 and 2. 
Fig.~\ref{hybrid} illustrates the control architecture of MLPIAC. At level 2 the control task of each area controller
is secondary frequency control for all nodes in the area
by centralized control. At level 1 the control task is
to reduce the overall control cost through coordination of the area marginal costs
by distributed control.

In the following two subsections, we define MLPIAC and analyze 
its properties during the transient phase and at the steady state.

\subsection{Definition of MLPIAC}  

\begin{definition}[MLPIAC]
Consider a large-scale power system (\ref{eq:system1}) with Assumption (\ref{assumption1}) and (\ref{assumption_comm}),
which is partitioned as in (\ref{partition}).
At level 2, the dynamic control law in area $A_r$ is described 
by the equations
\begin{subequations}\label{multi_level}
 \begin{align}
  \dot{\eta}_r&=\sum_{i\in\mathcal{V}_{M_r}\cup\mathcal{V}_{F_r}}D_i\omega_i
  +k_3 v_r,\label{Multi_eq1}\\
  \dot{\xi}_r&=-k_1\Big(\sum_{i\in\mathcal{V}_{M_r}}M_i\omega_i+\eta_r\Big)-k_2\xi_r,\label{Multi_eq2}\\
  0&=\frac{\lambda_r}{\alpha_r}-k_2\xi_r,\label{Multi_eq3}\\
  u_i&=\frac{\lambda_r}{\alpha_i}, ~i\in\mathcal{V}_{K_r}, \label{Multi_eq4}\\
  \frac{1}{\alpha_r}&=\sum_{i\in\mathcal{V}_{K_r}}{\frac{1}{\alpha_i}}
 \end{align}
\end{subequations}
where $\eta_r, \xi_r$ are state variables of the area controller, $v_r$ is an 
algebraic variable, $k_1, k_2, k_3\in (0,\infty)$ are parameters, $\alpha_r$ is a
constant defined as the control price of area $A_r$, $u_i$ is the control input at node $i$
in this area. 
At level 1, the coordination for the area is an algebraic equation only 
\begin{eqnarray}
  v_r=\sum_{q\in Z_m}{l_{rq}(\lambda_r(t)-\lambda_q(t))}\label{exchange}
\end{eqnarray}
 where $\{\lambda_r(t),~r\in Z_m\}$ are inputs and $\{v_r(t),~r\in Z_m\}$
 are outputs of the coordinators,
 $l_{rq}\in[0,\infty)$ is the weight of the communication line connecting area $A_r$ and $A_q$,
which can be chosen to accelerate the consensus of the marginal costs. 
 $l_{rq}$ defines a weighted undirected communication network with 
Laplacian matrix $(L_{rq})\in\mathbb{R}^{m\times m}$
\begin{eqnarray}\label{eq:laplacian2}
 L_{rq}=\begin{cases}
         -l_{rq}, &r\neq q,\\
         \sum_{k\neq i}l_{rk}, &r=q,
        \end{cases}
\end{eqnarray}
\end{definition}

In MLPIAC, at level 2, within area $A_r$, the area controller collects the local frequency 
deviations $\{\omega_i,~i\in\mathcal{V}_{M_r}\cup\mathcal{V}_{F_r}\}$ and calculates
the area control input $k_2\xi_r=\sum_{i\in\mathcal{V}_{K_r}}u_i$ by (\ref{Multi_eq1}, \ref{Multi_eq2}), the marginal cost of 
the area $\lambda_r$ by (\ref{Multi_eq3}) and the local control input $\{u_i,~i\in\mathcal{V}_{K_r}\}$
by (\ref{Multi_eq4}). The total control input
of area $A_r$ is denoted by $u_r=k_2\xi_r$.

In MLPIAC, at level 1, the area controllers exchange the marginal costs in order to achieve a consensus, which is a necessary condition for the global economic power dispatch as stated in 
(\ref{eq:kkt}).


MLPIAC has two special cases. (1) If each area consists of a single node, MLPIAC reduces
to a distributed control law, which is named \emph{Distributed Power-Imbalance Allocation Control} (DPIAC)
and has the following form
\begin{subequations}\label{eq:DPIAC}
 \begin{align}
\dot{\eta}_i&=D_i\omega_i+k_3\sum_{j\in\mathcal{V}}{l_{ij}{(k_2\alpha_i\xi_i-k_2\alpha_j\xi_j)}},\label{eq:DPIAC1}\\
\dot{\xi}_i&=-k_1(M_i\omega_i+\eta_i)-k_2\xi_i,\label{eq:DPIAC2}\\
u_i&=k_2\xi_i. \label{eq:DPIAC2u}
 \end{align}
\end{subequations}
(2) If the entire network is controlled as a single area, then it reduces to a centralized control law named 
\emph{Gather-Broadcast Power-Imbalance Allocation Control} (GBPIAC), which is described by 
\begin{subequations}\label{eq:controlPIAC3}
 \begin{align}
  \dot{\eta}_s&=\sum_{i\in\mathcal{V}_F}D_i\omega_i,\label{eq:controlPIAC3a}\\
  \dot{\xi}_s&=-k_1(\sum_{i\in\mathcal{V}_M}M_i\omega_i+\eta_s)-k_2\xi_s,\label{eq:controlPIAC3b}\\
   u_i&=\frac{\alpha_s}{\alpha_i}k_2\xi_s,~i\in\mathcal{V}_K,\label{eq:controlPIAC3c}\\
   \frac{1}{\alpha_s}&=\sum_{i\in\mathcal{V}_K}\frac{1}{\alpha_i}, \label{eq:controlPIAC3d}
 \end{align}
\end{subequations}
where $\eta_s,~\xi_s$ are state variables of the central controller.  
In the next subsection, it will be shown that MLPIAC includes a trade-off between DPIAC and GBPIAC.

\subsection{Properties of MLPIAC}
In this section, we analyse the properties of MLPIAC 
in the transient phase and steady state and verify whether it solves Problem \ref{problem:paper}. 
The following proposition describes the properties of MLPIAC during the transient phase. 
\begin{proposition}\label{theoremMPIAC}
Consider a large-scale power system partitioned as in (\ref{partition}), MLPIAC has the following properties
during the transient phase,
 \begin{enumerate}[(a)]
  \item at any time $t\in T$, the total control input $u_s$ of the system satisfies, 
\begin{subequations}\label{eq:appndx}
 \begin{align}
  \dot{\upsilon}_s&=P_s+u_s,\label{eq:appndxa}\\
 \dot{u}_s&=-k_2\big(k_1\upsilon_s+u_s\big),\label{eq:appndxb}
 \end{align}
\end{subequations}
where $\upsilon_s$ is an auxiliary variable. Furthermore, 
$u_s$ converges to $-P_s$ directly without any extra oscillations if $k_2\geq 4k_1$;
  \item at any time $t\in T$, within an area $A_r$, the inputs $\{u_i,~i\in\mathcal{V}_{K_r}\}$
  of the local controllers solve the following economic power dispatch problem 
  \begin{eqnarray}
 &&~\min_{\{u_i\in \mathbb{R},i\in\mathcal{V}_{K_r}\}} \sum_{i\in\mathcal{V}_{K_r} }\frac{1}{2}\alpha_iu_i^2
 \label{eq:optimal3}\\
 \nonumber
 &&s.t. ~~-u_r(t)+\sum_{i\in\mathcal{V}_{K_r} }{u_i(t)}=0. 
\end{eqnarray}
 \end{enumerate}
\end{proposition}

\emph{Proof:}
(a) Because of the symmetry of the matrix $(L_{rq})$, we derive from (\ref{exchange}) that, 
\begin{eqnarray}
 \sum_{r\in Z_m}v_r=\sum_{r, q\in Z_m}l_{rq}(\lambda_r-\lambda_q)=0. \label{eq:LRS}
\end{eqnarray}
 Summing all the equations in (\ref{eq:system1}), we obtain 
\begin{eqnarray}
  \sum_{i\in\mathcal{V}_M}M_i\dot{\omega}_i&=P_s-\sum_{i\in\mathcal{V}_M\cup\mathcal{V}_F}{D_i\omega_i}+u_s. \label{eq:sum}
\end{eqnarray}
Summing all the control inputs $\{u_r,~r\in Z_m\}$ with $u_r=k_2\xi_r$, we derive 
the total control input of the system as $u_s=\sum_{r\in Z_m}k_2\xi_r$. It follows from (\ref{eq:LRS}) and 
(\ref{Multi_eq2}) that 
\begin{eqnarray}
  \dot{u}_s(t)=-k_2\Big(k_1(\sum_{i\in\mathcal{V}_M}M_i\omega_i+\eta_s)+u_s(t)\Big),\label{eq:u_sum}
\end{eqnarray}
where $\eta_s=\sum_{r\in Z_m} \eta_r$ and following (\ref{Multi_eq1}) with derivative 
\begin{eqnarray}
 \dot{\eta}_s=\sum_{i\in\mathcal{V}_M\cup\mathcal{V}_F}D_i\omega_i \label{eq:eta_s}
\end{eqnarray}
Let $\upsilon_s(t)=\sum_{i\in\mathcal{V}_M}M_i\omega_i+\eta_s$, we derive (\ref{eq:appndxa})
 from (\ref{eq:sum}) and (\ref{eq:eta_s}), and (\ref{eq:appndxb}) from (\ref{eq:u_sum}). 
 Hence $u_s$ satisfies (\ref{eq:appndx}). The eigenvalues of the system (\ref{eq:appndx})
 are 
\begin{eqnarray}\label{eq:eigen}
 \mu=\frac{-k_2\pm\sqrt{k_2^2-4k_1k_2}}{2}. 
\end{eqnarray}
To avoid the extra oscillation caused by the oscillations of $u_s$, the second-order 
system (\ref{eq:appndx}) should be over-damped or critical-damped, 
thus the eigenvalues in (\ref{eq:eigen}) must be real. This needs $k_2\geq 4k_1$.
Hence if $k_2\geq 4k_1$, $u_s$ converges to $-P_s$ directly without any extra oscillations. 
 \\
 (b) Following (\ref{Multi_eq4}), we derive that at any time $t\in T$,
\begin{eqnarray*}
 \alpha_i u_i=\alpha_j u_j=\alpha_r k_2\xi_r, i,j\in\mathcal{V}_{K_r}.
\end{eqnarray*}
So the necessary condition (\ref{eq:kkt}) for the optimization problem (\ref{eq:optimal3}) 
is satisfied. Furthermore, with $\sum_{i\in\mathcal{V}_{K_r}}u_i(t)=u_r(t)$,
the optimization problem (\ref{eq:optimal3}) is solved at any time $t$. 
\hfill $\Box$

\begin{remark}
We use $u_s$ to estimate the power-imbalance $-P_s$ in system (\ref{eq:appndx}) which 
can be seen as an observer of $-P_s$. Similar to the high gain observer \cite{Khalil}, there may be overshoot in the initialization of the controllers due to the initial condition 
of the state. To the best of our knowledge, there do not seem general sufficient conditions on the system which guarantee that for all initial conditions the behavior of every state component is free of zero crossings and further eliminate this kind of overshoot. In this paper, we treat the 
case where the frequency trajectory fluctuates regularly and analyse the overshoots of $u_s$
caused by those fluctuations.  
\end{remark}

The marginal costs $\{\lambda_r,A_r\subset A\}$ of the areas are different during the transient phase, 
which will achieve a consensus due to the principle of consensus control. The consensus
speed of these marginal costs and the convergence speed of $u_s$ determine convergence of the control cost to its optimal solution. Because one objective 
of Problem \ref{problem:paper} is to eliminate the extra frequency oscillation, we set $k_2\geq 4k_1$ in this paper. 

In order to further investigate how MLPIAC improves the transient performance 
of the frequency and marginal cost, we decompose the dynamics of the 
power system into the following four subprocesses. 
\begin{enumerate}[(i)]
\item the convergence of $u_s$ to $-P_s$ as in (\ref{eq:appndx}) with a speed determined by $k_1$,
\item the synchronization of the frequency deviation $\omega_i(t)$ to $\omega_s(t)$ which 
is a physical characteristic of the power system (\ref{eq:system1}),
and the synchronization speed is determined by $\{u_i(t),~i\in\mathcal{V_{K}}\}$ and $\{D_i,~i\in\mathcal{V}_M\cup\mathcal{V}_F\}$;
\item the convergence of $\omega_{s}$ to $\omega_{syn}$ which further converges to zero as $u_s(t)$ converges to $-P_s$. This can be directly obtained from (\ref{eq:sync});
\item the consensus of the marginal costs $\{\lambda_r,A_r\subset A\}$ with a consensus speed 
determined by $k_3$ and $(L_{ij})$.  
\end{enumerate}

In GBPIAC, because the economic power dispatch problem is 
solved in a centralized way, the marginal costs of all nodes are identical.  

In MLPIAC, the transient performance of the frequency oscillation and marginal costs 
can be improved by tuning the corresponding coefficients of the four subprocesses.
The convergence of $\omega_{syn}$ to zero can be accelerated by a fast convergence of $u_s$, which 
can be obtained with a large $k_1$. The synchronization process of $\omega_i$ can be improved 
by tuning $\{D_i,~i\in\mathcal{V}_K\}$ which is the task of primary control. The corresponding 
actuators includes \emph{Power System Stabilizer} (PSS) systems embedded in the damping systems of the synchronous machines \cite{Milano2008}. The parameter
$D_i$ can be set as in \cite{Dorfler2014} or \cite{Motter2013} focusing 
on the control cost optimality and the small signal stability respectively.  
The convergence of $\{u_i,~i\in\mathcal{V}_K\}$ to their optimal solution can be 
improved by tuning $k_3$ or by a well designed $(L_{rq})$.

Furthermore, for a system with multiple areas, the size of the communication network with Laplacian matrix $(L_{rq})$ is much 
smaller than the one in DPIAC, the speed of achieving a consensus
marginal cost of the areas is much faster in MLPIAC, and the number of nodes in each 
area is smaller than the total number of nodes, the communication and computations of 
each area controller is reduced.


Before introducing Proposition \ref{proposition_equilibrium} which describes the properties of the steady state of the power system controlled by MLPIAC, 
we introduce the closed-loop system as follows. 
{\small
\begin{subequations}\label{eq:closed}
\begin{align}
  \hspace{-10pt}
  \dot{\theta}_i&=\omega_i,i\in\mathcal{V}_M\cup\mathcal{V}_F,\\
M_{i}\dot{\omega}_{i}&=P_{i}-D_{i}\omega_{i}-\sum_{j\in \mathcal V}{B_{ij}\sin{\theta_{ij}}}+\frac{\alpha_r}{\alpha_i} k_2\xi_r,i\in \mathcal{V}_{M_r}\subset \mathcal{V}_{M},\\
0&=P_{i}-D_{i}\omega_{i}-\sum_{j\in \mathcal V}{B_{ij}\sin{\theta_{ij}}}+\frac{\alpha_r}{\alpha_i} k_2\xi_r, i\in \mathcal{V}_{Fr}\subset \mathcal{V}_{F},\\
0&=P_{i}-\sum_{j\in \mathcal V}{B_{ij}\sin{\theta_{ij}}}, i\in \mathcal{V}_P, \\
\dot{\eta}_r&=\sum_{i\in\mathcal{V}_{M_r}\cup\mathcal{V}_{F_r}}D_i\omega_i
  +k_2k_3\sum_{q\in Z_m}{l_{rq}}(\alpha_r\xi_r-\alpha_q\xi_q),r\in Z_m,\label{eq:closed5}\\
  \dot{\xi}_r&=-k_1(\sum_{i\in\mathcal{V}_{M_r}}M_i\omega_i+\eta_r)-k_2\xi_r,~r\in Z_m\label{eq:closed6}
 \end{align}
\end{subequations}
}
where $\theta_{ij}=\theta_i-\theta_j$ for $(i,j)\in\mathcal{E}$, 
$i$ and $j$ are the node indices, and $r,q$ are the area indices. 

As in the Kuramoto-model \cite{Dorfler20141539}, the closed-loop system (\ref{eq:closed})
may not have a synchronous state 
if the power injections $\{P_i,\in\mathcal{V}\}$ are much larger that the line capacity 
$\{B_{ij},(i,j)\in\mathcal{E}\}$ \cite{Dorfler2014} and \cite{Xi2016}. 
We assume there exists a synchronous state for the power system, which 
can be satisfied by reserving some margin in the line capacity by tertiary control.
\begin{assumption}\label{assumption_equlibrium}
 There exists a synchronous state for the closed-loop system (\ref{eq:closed}) such that 
 $$\theta^*\in\Theta=\Big\{\theta_i\in\mathbb{R},~\forall i\in\mathcal{V}\big||\theta_i-\theta_j|<\frac{\pi}{2}, 
 \forall(i,j)\in\mathcal{E}\Big\}$$
 where $\theta^*=\text{col}(\theta_i^*)\in\mathbb{R}^{n_t}$, $n_t$ is the total number of nodes in $\mathcal{V}$. 
 The condition $\theta^*\in\Theta$ is commonly referred to as a security constraint \cite{DePersis2016} 
 and restricts the equilibrium to desired power flows. 
\end{assumption}
~~Note that the equilibria with $\theta$ out of $\Theta$ usually leads to undesired power flows 
which either have cyclic power flows or are unstable \cite{Xi2016}.

For the synchronous state of the closed-loop system, 
we have the following proposition.
\begin{proposition}\label{proposition_equilibrium}
 If the assumptions (\ref{assumption1}), (\ref{assumption_comm}) and (\ref{assumption_equlibrium}) hold, 
 then there exists at most one synchronous state for the closed-loop system (\ref{eq:closed})
 such that
    \begin{subequations}
 \begin{align}
    \theta_i^*&\in\Theta,~i\in\mathcal{V}\\
   \omega^*_i&=0,\mathcal{V}_M\cup\mathcal{V}_F,\label{eq:equilibrum1}\\
  P_s+k_2\sum_{r\in Z_m}{\xi_r^*}&=0, \label{eq:equilibrum2}\\
  k_1\eta_r^*+k_2{\xi_r^*}&=0, ~r\in Z_m,\label{eq:equilibrum3}\\
  \alpha_r\xi_r^*-\alpha_q\xi_q^*&=0, ~\forall ~r,q\in Z_m, \label{eq:equilibrum4}\\
  \alpha_i u_i^*-k_2\alpha_r\xi_r^*&=0,~ i\in\mathcal{V}_{K_{r}}\subset \mathcal{V}_{K}.\label{eq:equilibrium5}
  \end{align}
 \end{subequations}
\end{proposition}
~~\emph{Proof:}  From Proposition \ref{theoremMPIAC}, the dynamics of $u_s$ satisfies (\ref{eq:appndx}),
which yields that $u_s^*=-P_s$ at the synchronous state. Thus (\ref{eq:equilibrum2}) is derived 
with $u_s(t)=\sum_{r\in Z_m}{k_2\xi_r}(t)$.
Following (\ref{eq:sync}), we further derive that $\omega_{syn}=0$,
which yields (\ref{eq:equilibrum1}) with the definition of the synchronous state (\ref{eq:synchronize}). 
By (\ref{eq:closed6}), $\omega_i^*=0$ and  $\dot{\xi}_i^*=0$ for all $i\in\mathcal{V}_K$, we derive 
(\ref{eq:equilibrum3}). 
By (\ref{eq:closed5}) and $\omega_i^*=0$, we obtain (\ref{eq:equilibrum4}). 
By (\ref{Multi_eq3}) and (\ref{Multi_eq4}), we arrive at (\ref{eq:equilibrium5}). 
From (\ref{eq:equilibrum4},\ref{eq:equilibrium5}) it follows that $\alpha_i u_i^*=\alpha_j u_j^*$ for all $i,j\in\mathcal{V}_K$,
thus the necessary condition (\ref{eq:kkt}) is satisfied. Following (\ref{eq:equilibrum2}), it yields 
$P+\sum_{i\in\mathcal{V}_K}u_i^*=0$ and the economic dispatch problem (\ref{eq:optimal1}) is solved subsequently. 
Accoding to \cite{Araposthatis1981,skar_uniqueness_equilibrium}, there exists at most one synchronous state such that  
$\theta^*\in\Theta$. \hfill $\Box$

With the improved transient performance as stated in the dynamics decomposition and
the properties of the system at the steady state as in Proposition \ref{proposition_equilibrium}, 
Problem \ref{problem:paper} is solved. 
\begin{remark}\label{MLPIACPI}
MLPIAC actually includes proportional and integral control input,
which are the terms $k_1\sum_{i\in\mathcal{V}_{M_r}}M_i\omega_i$ and 
$k_1\eta_r$ respectively in (\ref{Multi_eq2}). In order 
to get a desired performance, the parameters $M_i$ and $D_i$ should be known. 
In practice, they are known for traditional 
synchronous machines \cite{Kundur1994}. However, they may not be known for the frequency 
dependent nodes. In that case, the uncertainties from these 
parameters can be added to the power-imbalance, which becomes a 
time-varying value and can be compensated by the controllers at the steady state because 
of the included integral control input. Theoretical analysis on the robustness of MLPIAC 
with these uncertainties needs to be further studied. 
\end{remark}

\begin{remark}
Through eigenvalue analysis or $\mathcal{H}_2/\mathcal{H}_\infty$ norm optimization, the 
existing control laws, e.g., the robust PI control in \cite{Hassan_Bevrani},
the distributed averaging PI control \cite{Dorfler2014,Andreasson2013}, and 
the distributed control law in \cite{XiangyuWu2017},
may also obtain the desired transient performance as that of MLPIAC.
However, intensive computations are needed and the mechanism to improve 
the transient performance is not as clear as in MLPIAC. In addition, the multi-level 
control structure of MLPIAC has not been considered in these methods. 
\end{remark}

%

\section{Stability analysis of MLPIAC}\label{Sec:stability}

In Proposition \ref{problem:paper}, we have proven that the total control 
input converges to the power-imbalance directly when $k_2\geq 4k_1$, 
which does not imply 
the state of the power system converges to its steady state. 
In this section, we focus on the asymptotic stability of MLPIAC.

The power flows $\{B_{ij}\sin{(\theta_{ij})},(i,j)\in\mathcal{E}\}$ only depend on the angle differences.
As in \cite{PIAC}, we choose a reference angle, e.g., $\theta_1\in\mathcal{V}_M$, and transform the 
system state into a new coordinate system such that 
\begin{eqnarray*}
 \varphi_i=\theta_i-\theta_1, ~i\in\mathcal{V}. 
\end{eqnarray*}
which yields $\dot{\varphi}_i=\omega_i-\omega_1$ for all $i$ in $\mathcal{V}_M\cup\mathcal{V}_F$. In the new coordinate, 
the closed-loop system (\ref{eq:closed}) becomes
{\small
\begin{subequations}\label{closed_MPIAC2}
\begin{align}
  \hspace{-6pt}
  \dot{\varphi}_i&=\omega_i-\omega_1,i\in\mathcal{V}_M\cup\mathcal{V}_F,\\
    \hspace{-6pt}
M_{i}\dot{\omega}_{i}&=P_{i}-D_{i}\omega_{i}-\sum_{j\in \mathcal V}{B_{ij}\sin{\varphi_{ij}}}+\frac{\alpha_r}{\alpha_i} k_2\xi_r,~i\in \mathcal{V}_{M},\\
  \hspace{-6pt}
D_i\dot{\varphi}_i&=P_{i}-D_{i}\omega_{1}-\sum_{j\in \mathcal V}{B_{ij}\sin{\varphi_{ij}}}+\frac{\alpha_r}{\alpha_i} k_2\xi_r,~i\in\mathcal{V}_{F},\\
  \hspace{-6pt}
0&=P_{i}-\sum_{j\in \mathcal V}{B_{ij}\sin{\varphi_{ij}}}, ~i\in \mathcal{V}_P, \label{eq:closed_DPIAC2d}\\
  \hspace{-6pt}
\dot{\eta}_r&=\sum_{i\in\mathcal{V}_{M_r}\cup\mathcal{V}_{F_r}}D_i\omega_i
  +k_2k_3\sum_{q\in Z_m}{l_{rq}}(\alpha_r\xi_r-\alpha_q\xi_q),~r\in Z_m,\\
   \hspace{-6pt}
 \dot{\xi}_r&=-k_1(\sum_{i\in\mathcal{V}_{M_r}}M_i\omega_i+\eta_r)-k_2\xi_r,~r\in Z_m
 \end{align}
\end{subequations}
}
which can be written in the form of DAEs (\ref{eq:DAEtt}) of Appendix \ref{appendix_DAE}.
Following Assumption (\ref{assumption_equlibrium}),
$\varphi=\text{col}(\varphi_i)\in\mathbb{R}^{n_t}$ satisfies
\begin{eqnarray*}
 \varphi\in\Phi=\big\{\varphi_i\in\mathbb{R},i\in\mathcal{V}||\varphi_i-\varphi_j|\leq \frac{\pi}{2},\forall(i,j)\in\mathcal{E},~\varphi_1=0\big\}.
\end{eqnarray*}

We make the following assumption that the control gain coefficients 
$k_1,k_2,k_3$ satisfy a certain condition (that will be proven as a sufficient condition of the asymptotic stability of MLPIAC).
\begin{assumption}\label{assumption_coefficient}
 Assume the control gain coefficients, $k_1, k_2, k_3$, satisfy that 
 \begin{eqnarray*}
  \frac{k_2}{k_1}>\frac{2(\alpha D)_{\max}}{(\alpha D)_{\min}(1+2k_3\lambda_{\min})}
 \end{eqnarray*}
\end{assumption}
where $(\alpha D)_{\min}=\min\{\alpha_i D_i,~i\in\mathcal{V}_K\}$, $(\alpha D)_{\max}=\max\{\alpha_i D_i,~i\in\mathcal{V}_K\}$ and 
$\lambda_{\min}$ is the smallest nonzero eigenvalue of matrix $L\alpha_R$ 
where $L=(L_{rq})\in \mathbb{R}^{m\times m}$ is defined in (\ref{eq:laplacian2}) and 
$\alpha_{R}=\text{diag}(\alpha_r)\in \mathbb{R}^{m\times m}$.

We rewrite the state and algebraic variables into a vector form, $(\varphi,\omega,\eta,\xi)\in\mathbb{R}^{n_t}\times\mathbb{R}^n\times
\mathbb{R}^{m}\times\mathbb{R}^m$. 
 The following theorem states the asymptotic stability of the equilibrium of MLPIAC. 
\begin{theorem}\label{theorem_stability}
If assumptions (\ref{assumption1},\ref{assumption_comm},\ref{assumption_equlibrium}) and (\ref{assumption_coefficient}) hold, 
then for the closed-loop system (\ref{eq:closed}), 
  \begin{enumerate}[(a)]
 \item there exists an unique synchronous state
 $z^*=(\varphi^*,\omega^*,\eta^*,\xi^*)\in\Psi$ where $\Psi=\Phi\times\mathbb{R}^{n}\times\mathbb{R}^{m}\times \mathbb{R}^{m}$. 
 \item there exists a domain $\Psi^d\subset\Psi$ such that 
 starting at any initial state $z^0=(\varphi^0,\omega^0,\eta^0,\xi^0)\in \Psi^d$ which satisfies
 the algebraic equations (\ref{eq:closed_DPIAC2d}), the state trajectory converges to the unique synchronous state $z^*\in\Psi$.
 \end{enumerate}
\end{theorem}


\begin{remark}
 In MLPIAC, Assumptions (\ref{assumption1}, \ref{assumption_comm}, \ref{assumption_equlibrium}) and 
 (\ref{assumption_coefficient}) are both necessary and realistic at the same time. Assumptions (\ref{assumption1}) and (\ref{assumption_comm})
 are necessary for the implementation of MLPIAC to solve the economic power dispatch problem (\ref{eq:optimal1}). Assumptions (\ref{assumption_equlibrium}) and (\ref{assumption_coefficient})
 are general sufficient conditions for the stability of MLPIAC. Assumption (\ref{assumption1}) and 
 (\ref{assumption_equlibrium}) can be guaranteed by tertiary control and Assumption (\ref{assumption_comm})
 by an effective communication infrastructure.
\end{remark}

\begin{remark}\label{Remark_coefficient}
The inverse of damping coefficient, $\frac{1}{D_i}$ can be viewed as the control cost of primary control
\cite{Dorfler2014}. When $\alpha=\gamma D^{-1}$, $\gamma\in\mathbb{R}$ is a positive number,  which indicates that the cost of 
the secondary frequency control is proportional to the primary control cost and leads to $(\alpha D)_{\min}=(\alpha D)_{\max}$, 
DPIAC is asymptotically stable if $k_2>2k_1$. Specially, Assumption (\ref{assumption_coefficient}) can be dropped
in GBPIAC in the theoretical analysis as will be explained in Remark (\ref{GBPIAC_assumption_relaxed}). For DPIAC and MLPIAC, 
our numerical simulations have shown 
that the control law is asymptotically stable though assumption (\ref{assumption_coefficient}) is not satisfied, which will be shown in section \ref{Sec:case}. 
\end{remark}

In the following, we will prove Theorem (\ref{theorem_stability}). 
The closed-loop system (\ref{closed_MPIAC2}) is rewritten in a vector form as follows,
\begin{subequations}\label{closed_MLPIAC_vector}
 \begin{align}
 \dot{\tilde{\varphi}}&=\omega-\omega_1 1_n,\\
  M\dot{\omega}&=P-D\omega-P^t+k_2\alpha^{-1}R\alpha_R\xi, \label{closed_MLPIAC_vb}\\
  0&=\tilde{P}-\tilde{P}^t,\label{closed_MLPIAC_vc}\\
  \dot{\eta}&=R^TD\omega+k_2k_3L\alpha_R\xi,\label{closed_MLPIAC_vd}\\
  \dot{\xi}&=-k_1(R^TM\omega+\eta)-k_2\xi, \label{closed_MLPIAC_ve}
 \end{align}
\end{subequations}
where $\tilde{\varphi}=\text{col}(\varphi_i)$ with $i\in\mathcal{V}_M\cup\mathcal{V}_F$, $P=\text{col}(P_i)\in\mathbb{R}^n$ for $i\in\mathcal{V}_M\cup\mathcal{V}_F$,
$\tilde{P}=\text{col}(P_j)\in\mathbb{R}^{n_p}$ for $j\in\mathcal{V}_P$, $P^t=\text{col}(P_i^t)\in\mathbb{R}^n$
with $P_i^t=\sum_{j\in\mathcal{V}}{B_{ij}\sin{\varphi_{ij}}}$ for $i\in\mathcal{V}_M\cup\mathcal{V}_F$,
$\tilde{P}^t=\text{col}(\tilde{P}_i^t)\in\mathbb{R}^{n_p}$
with $\tilde{P}_i^t=\sum_{j\in\mathcal{V}}{B_{ij}\sin{\varphi_{ij}}}$ for $i\in\mathcal{V}_P$. Note 
that $\tilde{\varphi}$ only includes the variables $\{\varphi_i,i\in\mathcal{V}_M\cup\mathcal{V}_F\}$ while $\varphi$ includes $\{\varphi_i,i\in\mathcal{V}\}$.  
For the definitions of $M, D, R,\alpha_R$, we refer to Appendix \ref{appendix_notation}. 

We transform the state of the control law (\ref{closed_MLPIAC_vd},\ref{closed_MLPIAC_ve}) to 
a new coordinate as a preparation for the stability analysis of MLPIAC. 
Following Theorem \ref{theorem:Symmetrizable} in Appendix 
\ref{appendixSymmetrizable}, let $\rho=Q^{-1}\eta$ and $\sigma=Q^{-1}\xi$. The vector form 
(\ref{closed_MLPIAC_vd},\ref{closed_MLPIAC_ve})
becomes
\begin{subequations}
 \begin{align*}
  \dot{\rho}&=Q^{-1}R^TD\omega+k_2k_3\Lambda \sigma,\\
  \dot{\sigma}&=-k_1Q^{-1}R^TM\omega-k_1\rho-k_2\sigma, 
 \end{align*}
\end{subequations}
where all the components $(\rho_i,~\sigma_i)$ of $(\rho,~\sigma)$ are decoupled from
each other. When writing the dynamics of $(\rho_i,~\sigma_i)$ separately, we derive 
\begin{subequations}\label{DPIAC_decomposed}
 \begin{align}
  \dot{\rho}_i&=Q_{vi}^TR^TD\omega+k_2k_3\lambda_i \sigma_i,\\
  \dot{\sigma}_i&=-k_1Q_{vi}^TR^TM\omega-k_1\rho_i-k_2\sigma_i, 
 \end{align}
\end{subequations}
where $Q^{-1}$ is decomposed into vectors, i.e., $Q^{-1}=(Q_{v1},Q_{v2},\cdots,Q_{vn})$. 
In (\ref{DPIAC_decomposed}), the controller of component $i$ calculates the output $\sigma_i$ with 
the input $\omega$ for the power system. We investigate the dynamic behavior of 
the component $(\rho_1,\sigma_1)$ and $\{(\rho_i,\sigma_i),~i=2,\cdots,n\}$ of $(\rho,\sigma)$
respectively. 

For the first component $(\rho_1,\sigma_1)$ of $(\rho,\sigma)$, we have the following lemma.
\begin{lemma}\label{lemma_component1}
 The dynamics of $(\rho_1,\sigma_1)$ described by (\ref{DPIAC_decomposed}) is identical 
 to that of $(\eta_s,\xi_s)$ in (\ref{eq:controlPIAC3}).
\end{lemma}
\emph{Proof:} 
By (\ref{DPIAC_decomposed}) and (\ref{eq:R2}), $\lambda_1=0$ and $Q_{vi}=1_m$ from (\ref{eq2:symmetrizable1}),
 we derive the dynamics of $(\rho_1,\sigma_1)$ as follows
\begin{subequations}\label{eq:dynamics_rho1}
\begin{align}
 \dot{\rho}_1&=1_mR^TD\omega=\sum_{i\in\mathcal{V}_M\cup\mathcal{V}_F}{D_i\omega_i},\\
 \dot{\sigma}_1&=-k_1(1_mR^TM\omega+\rho_1)-k_2\sigma_1\\
 &=-k_1(\sum_{i\in\mathcal{V}_M}{M_i\omega_i}+\rho_1)-k_2\sigma_1.
\end{align}
\end{subequations}
 In addition,
by summing all the equations in
(\ref{closed_MLPIAC_vb}) for all $i\in\mathcal{V}$, we derive
\begin{eqnarray}
  \sum_{i\in\mathcal{V}_M}M_i\dot{\omega}_i&&=P_s-\sum_{i\in\mathcal{V}_M\cup\mathcal{V}_F}{D_i\omega_i}+k_21_n^T\alpha^{-1}R\alpha_R\xi\nonumber\\
  &&~~~~\text{by (\ref{eq:R})}\nonumber\\
  &&=P_s-\sum_{i\in\mathcal{V}_M\cup\mathcal{V}_F}{D_i\omega_i}+k_21_m^T\xi\nonumber\\
  &&~~~~\text{by (\ref{eq:symmetraizable1})}\nonumber\\
  &&=P_s-\sum_{i\in\mathcal{V}_M\cup\mathcal{V}_F}{D_i\omega_i}+k_2\sigma_1.\label{summing}
\end{eqnarray}
So $k_2\sigma_1$ is the control input for the power system (\ref{eq:system1}) as $k_2\xi_s$. 
Furthermore, the initial values of $(\rho_1,\sigma_1)$ and $(\eta_s,\xi_s)$ are identical, which 
are both computed from $\{\omega_i(0),i\in\mathcal{V}_K\}$, so
 the dynamics of $(\rho_1,\sigma_1)$ is identical to that of $(\eta_s,\xi_s)$ in (\ref{eq:controlPIAC3})
if under the same initial values. \hfill $\Box$

As described in Remark \ref{MLPIACPI}, the total control ipnut of 
MLPIAC includes a proportional and an integral control input. With the superposition principle, we decompose the dynamics of $(\rho_i,\sigma_i)$ for $i=2,\cdots,m$ for the 
proportional and the integral input into the following two independent dynamics. 
\begin{subequations}
 \begin{align*}
  \dot{\rho}_{mi}&=k_2k_3\lambda_i\sigma_{mi},\\
  \dot{\sigma}_{mi}&=-k_1Q_{vi}^TR^TM\omega-k_1\rho_{mi}-k_2\sigma_{mi},
 \end{align*}
\end{subequations}
and 
\begin{subequations}
 \begin{align*}
   \dot{\rho}_{di}&=Q_{vi}^TR^TD\omega+k_2k_3\lambda_i\sigma_{di},\\
  \dot{\sigma}_{di}&=-k_1\rho_{di}-k_2\sigma_{di},
 \end{align*}
\end{subequations}
from which it can be easily derived that $\rho_i=\rho_{mi}+\rho_{di}$ and 
$\sigma_i=\sigma_{mi}+\sigma_{di}$. 

In the coordinate of $(\varphi,\omega,\rho_1,\sigma_1,\rho_m,\sigma_m,\rho_d,\sigma_d)$, 
the closed-loop system (\ref{closed_MLPIAC_vector}) becomes
\begin{subequations}\label{closed_MLPIAC_vector2}
 \begin{align}
 \dot{\tilde{\varphi}}&=\omega-\omega_1 1_n,\\
  M\dot{\omega}&=P-D\omega-P^t+k_2\alpha^{-1}R\alpha_RQ\sigma,\\
  0&=\tilde{P}-\tilde{P}^t,\\
\dot{\rho}_1&=1_m^TR^TD\omega,\\
\dot{\sigma}_1&=-k_1 1_m^TR^TM\omega-k_1\rho_1-k_2\sigma_1,\\
\dot{\rho}_{m}&=k_2k_3\Lambda\sigma_{m},\\
\dot{\sigma}_{m}&=-k_1W^TR^TM\omega-k_1\rho_{m}-k_2\sigma_{m},\\
\dot{\rho}_{d}&=W^TR^TD\omega+k_2k_3\Lambda\sigma_{d},\\
\dot{\sigma}_{d}&=-k_1\rho_{d}-k_2\sigma_{d},
 \end{align}
\end{subequations}
where $\sigma=\text{col}(\rho_i)$, $\rho_i=\rho_{mi}+\rho_{di}$ for $i=2,\cdots,m$, $W$ is defined 
as in Theorem (\ref{theorem:Symmetrizable}). Note that $\Lambda\in\mathbb{R}^{(m-1)\times(m-1)}$ only
includes the nonzero eigenvalues of $L\alpha_R$, which is different from the one
in Appendix \ref{appendix_notation}.

Following Proposition (\ref{proposition_equilibrium}), for the closed-loop system (\ref{closed_MLPIAC_vector2})
we have the following Lemma on the equilibrium state. 
\begin{lemma}\label{lemma_equilibrium_vector}
In the coordinate of $(\varphi, \omega, \rho_1,\sigma_1,\rho_m,\sigma_m,\rho_d,\sigma_d)$, the unique equilibrium state $(\theta^*,\omega^*,\eta^*,\xi^*)$ 
of the closed-loop system (\ref{eq:closed}) proposed in 
Proposition (\ref{proposition_equilibrium}) is equivalent to $(\varphi^*,\omega^*,\rho^*_1,\sigma_1^*,\rho^*_m,\sigma^*_m,\rho^*_d,\sigma^*_d)
\in\Phi\times\mathbb{R}^n\times\mathbb{R}\times\mathbb{R}\times\mathbb{R}^{m-1}\times\mathbb{R}^{m-1}
\times\mathbb{R}^{m-1}\times\mathbb{R}^{m-1}$ such 
that 
\begin{subequations}
 \begin{align}
  \varphi^*\in\Phi&=\big\{\varphi\in \mathbb{R}^{n_t}||\varphi_i-\varphi_j|<\frac{\pi}{2},~\forall (i,j)\in\mathcal{E}, \varphi_1=0\big\},\\
  \omega_i^*&=0,~i\in\mathcal{V}_M\cup\mathcal{V}_F,\\
  k_1\rho_1^*+k_2\sigma^*_1&=0,\label{equilibrium2c}\\
  k_2\sigma^*_1+P_s&=0,\label{equilibrium2d}\\
  \rho_{mi}^*&=0,~i=2,\cdots,n,\label{equilibrium2e}\\
  \sigma_{mi}^*&=0,~i=2,\cdots,n,\label{equilibrium2f}\\
  \rho_{di}^*&=0,~i=2,\cdots,n,\label{equilibrium2g}\\
  \sigma_{di}^*&=0,~i=2,\cdots,n,\label{equilibrium2h}
 \end{align}
\end{subequations}
\end{lemma}
\emph{Proof:}
When mapping $\theta$ to $\varphi$, we can easily 
obtain 
$\varphi\in\Phi=\big\{\varphi\in \mathbb{R}^{n_t}||\varphi_i-\varphi_j|<\frac{\pi}{2},~\forall (i,j)\in\mathcal{E}, \varphi_1=0\big\}$,
$\omega^*=0$ can be directly derived from Proposition (\ref{proposition_equilibrium}). 
 
By Lemma (\ref{lemma_component1}), we have  $(\rho_1^*,\sigma_1^*)=(\eta_s^*,\sigma_s^*)$ at the steady state, 
which yields (\ref{equilibrium2c}) and (\ref{equilibrium2d}).

By the dynamics (\ref{DPIAC_decomposed}) of $(\rho_{mi},\sigma_{mi})$ and that of 
$(\rho_{di},\sigma_{di})$, we derive that $(\rho_{mi}^*,\sigma_{mi}^*)=(0,0)$ and $(\rho_{di}^*,\sigma_{di}^*)=(0,0)$ for all $i=2,\cdots,n$, at 
the steady state, which lead to  (\ref{equilibrium2e})-(\ref{equilibrium2h}). 
\hfill $\Box$

In order to prove the asymptotic stability of the equilibrium $(\theta^*,\omega^*,\eta^*,\xi^*)$,
we only need to prove the asymptotic stability of the equilibrium 
$(\varphi^*,\omega^*,\rho^*_1,\sigma_1^*,\rho^*_m,\sigma^*_m,\rho^*_d,\sigma^*_d)$.
We define function 
\begin{eqnarray}
U(\varphi)=\sum_{(i,j)\in\mathcal{E}}B_{ij}(1-\cos{(\varphi_i-\varphi_j)})
\end{eqnarray}
and variable $\upsilon_s=\sum_{i\in\mathcal{V}_M}{M_i\omega_i}+\rho_1$. By (\ref{summing})  and (\ref{eq:dynamics_rho1}), we obtain
dynamics of $(\upsilon_s,\sigma_1)$,
\begin{subequations}\label{dynamcis_upsilon}
 \begin{align}
  \dot{\upsilon}_s&=P_s+k_2\sigma_1,\\
  \dot{\sigma}_1&=-k_1\upsilon_s-k_2\sigma_1,
 \end{align}
\end{subequations}
with equilibrium state $(\upsilon^*_s,\sigma^*_1)=(\frac{1}{k_1}P_s,-\frac{1}{k_2}P_s)$.

In the following, we prove the equilibrium  $(\theta^*,\omega^*,\rho^*_1,\sigma_1^*,\rho^*_m,\sigma^*_m,\rho^*_d,\sigma^*_d)$.
is locally asymptotically stable following Lyapunov method. 

\emph{Proof of Theorem (\ref{theorem_stability}):}
The existence and uniqueness of the synchronous state in $\Psi$ follows Proposition (\ref{proposition_equilibrium}) directly.
Since the closed-loop system (\ref{eq:closed}) is equivalent to (\ref{closed_MLPIAC_vector2}), we prove the 
equilibrium $(\varphi^*,\omega^*,\rho^*_1,\sigma_1^*,\rho^*_m,\sigma^*_m,\rho^*_d,\sigma^*_d)$ of (\ref{closed_MLPIAC_vector2})
is locally asymptotically stable. 
The proof follows Theorem (\ref{thm1}). It follows \cite[Lemma 5.2]{PIAC} that the algebraic equations (\ref{eq:closed_DPIAC2d}) are regular.
In addition, there exists
an unique equilibrium for the closed-loop system (\ref{closed_MLPIAC_vector2}) following Lemma (\ref{lemma_equilibrium_vector}),
we only need to find a Lyapunov function $V(x,y)$ as in Theorem (\ref{thm1}).

Before introducing the Lyapunov function candidate, we define the following functions.
\begin{subequations}
 \begin{align*}
 V_0&=U(\varphi)-U(\varphi^*)-\nabla_{\varphi}U(\varphi^*)(\varphi-\varphi^*)+\frac{1}{2}\omega^TM\omega,\\
 V_1&=(c_1+1)\Big(\frac{1}{2k_1}(k_1\upsilon_s-k_1\upsilon^*_s)^2+\frac{1}{2k_2}(k_2\sigma_1-k_2\sigma_1^*)^2\Big)\\
 &~~+\frac{1}{2k_2}(k_2\sigma_1-k_2\sigma^*_1)^2+\frac{1}{2k_1}(k_1\upsilon_s+k_2\sigma_1)^2,
 \end{align*}
 \end{subequations}
where $c_1\in\mathbb{R}$ and $k_1\upsilon_s^*+k_2\sigma_1^*=0$ has been used. Denote $x_m=k_1\rho_m$, $y_m=k_2\sigma_m$, $z_m=k_1\rho_m+k_2\sigma_m$,  $x_d=k_1\rho_d$, $y_d=k_2\sigma_d$,
$z_d=k_1\rho_d+k_2\sigma_d$ and define
 \begin{subequations}
 \begin{align*}
 V_m&=\frac{\beta_m }{2}x_m^TC_m x_m+\frac{(1+\beta_m)k_1k_3}{2k_2}y_m^TC_m\Lambda y_m+\frac{1}{2}z_m^TC_m z_m,\\
 V_d&=\frac{\beta_dc_d}{2}x_d^Tx_d+\frac{(1+\beta_d)c_dk_1k_3}{2k_2}y_d^T\Lambda y_d+\frac{c_d}{2}z_d^Tz_d,
 \end{align*}
\end{subequations}
where  $\beta_m\in\mathbb{R}$, $\beta_d\in\mathbb{R}$, $c_d\in\mathbb{R}$ are positive 
and $C_m=\text{diag}(c_{mi})\in\mathbb{R}^{(m-1)\times (m-1)}$ with all the diagonal elements $c_{mi}>0$, 
and $\Lambda\in\mathbb{R}^{(m-1)\times(m-1)}$ is a diagonal matrix with diagonal elements 
being the nonzero eigenvalues of $L\alpha_R$.  

In the following, we focus on the derivatives of the functions $V_0, V_1, V_m, V_d$. The derivative of $V_0$ is 
\begin{subequations}
 \begin{align*}
\dot{V}_0&=-\omega^TD\omega+(k_2\sigma-k_2\sigma^*)^TQ^T\alpha_RR^T\alpha^{-1}\omega\\
  &{\scriptsize~~\text{by~$Q=[Q_1,Q_2,\cdots,Q_n]$ and $S=[Q_2,\cdots,Q_n]$.}}\\
  &=-\omega^TD\omega+(k_2\sigma_1-k_2\sigma^*_1)Q_1^T\alpha_RR^T\alpha^{-1}\omega\\
  &~~+(k_2\sigma-k_2\sigma^*)^TS^T\alpha_RR^T\alpha^{-1}\omega\\
  &=-\omega^TD\omega+(k_2\sigma_1-k_2\sigma^*_1)Q_1^T\alpha_RR^T\alpha^{-1}\omega\\
  &~~+(k_2\sigma_m-k_2\sigma^*_m)^TS^T\alpha_RR^T\alpha^{-1}\omega\\
  &~~+(k_2\sigma_d-k_2\sigma^*_d)^TS^T\alpha_RR^T\alpha^{-1}\omega,\\
  &~~\text{\footnotesize by $S^T\alpha_R=W^T$, $Q_1^T\alpha_R=Q_{v_1}^T$, $\sigma_m^*=0$}\\
  &=-\omega^TD\omega+(k_2\sigma_1-k_2\sigma^*_1)Q_{v_1}^TR^T\alpha^{-1}\omega\\
  &~~+(k_2\sigma_m)^TW^TR^T\alpha^{-1}\omega+(k_2\sigma_d)^TW^TR^T\alpha^{-1}\omega\\
  &=-\omega^TD\omega+(k_2\sigma_1-k_2\sigma^*_1)Q_{v_1}^TR^T\alpha^{-1}\omega\\
  &~~+y_m^TW^TR^T\alpha^{-1}\omega+y_d^TW^TR^T\alpha^{-1}\omega.
   \end{align*}
\end{subequations}  
The derivative of $V_1$ can be rewritten by adding and subtracting a term
\begin{subequations}
 \begin{align*}
\dot{V}_1&=-(c_1+1)(k_2\sigma_1-k_2\sigma_1^*)^2-\frac{k_2}{k_1}(k_1\upsilon+k_2\sigma_1)^2\\
  &=-(k_2\sigma_1-k_2\sigma^*_1)Q_{v1}^TR^T\alpha^{-1}\omega\\
  &~~+(k_2\sigma_1-k_2\sigma^*_1)Q_{v1}^TR^T\alpha^{-1}\omega\\
  &~~-(c_1+1)(k_2\sigma_1-k_2\sigma_1^*)^2-\frac{k_2}{k_1}(k_1\upsilon+k_2\sigma_1)^2.
   \end{align*}
\end{subequations}
By inequalities 
\begin{subequations}
 \begin{align*}
  &(k_2\sigma_1-k_2\sigma^*_1)Q_{v1}^TR^T\alpha^{-1}\omega\leq \frac{1}{8}\omega^T\epsilon D\omega\\
  &~~~~~~~~~~~+2(k_2\sigma_1-k_2\sigma_1^*)^2 Q_{v1}^TR^T\alpha^{-2}(\epsilon D)^{-1}RQ_{v1},
 \end{align*}
\end{subequations}
we obtain
  \begin{subequations}
 \begin{align*}
 \dot{V}_1&\leq -(k_2\sigma_1-k_2\sigma^*_1)Q_{v1}^TR^T\alpha^{-1}\omega+\frac{1}{8}\omega^T\epsilon D\omega\\
  &~~+2(k_2\sigma_1-k_2\sigma_1^*)^2 Q_{v1}^TR^T\alpha^{-2}(\epsilon D)^{-1}RQ_{v1}\\
  &~~-(c_1+1)(k_2\sigma_1-k_2\sigma_1^*)^2-\frac{k_2}{k_1}(k_1\upsilon_s+k_2\sigma_1)^2\\
  &~~\text{\small let $c_1=2Q_{v1}^TR^T\alpha^{-2}(\epsilon D)^{-1}RQ_{v1}$}\\
  &=-(k_2\sigma_1-k_2\sigma^*_1)Q_{v1}^TR^T\alpha^{-1}\omega+\frac{1}{8}\omega^T\epsilon D\omega\\
  &~~-(k_2\sigma_1-k_2\sigma_1^*)^2-\frac{k_2}{k_1}(k_1\upsilon+k_2\sigma_1)^2.\\
 \end{align*}
\end{subequations}
The derivative of $V_m$ is 
{\small
\begin{subequations}
 \begin{align*}
 \dot{V}_m&=-k_2z_m^TC_m z_m-\beta_mk_1k_3y_m^TC_m\Lambda y_m\\
 &~~-k_1k_2 z_m^TC_mW^TR^TM\omega\\
 &~~-(1+\beta_m)k_1^2k_3y_m^TC_m\Lambda W^TR^TM\omega\\
 &=-y_m^TW^TR^T\alpha^{-1}\omega+y_m^TW^TR^T\alpha^{-1}\omega-k_2 z_m^TC_m z_m\\
 &~~-\beta_mk_1k_3y_m^TC_m\Lambda y_m-k_1k_2z_m^TC_mW^TR^TM\omega\\
 &~~-(1+\beta_m)k_1^2k_3y_m^TC_m\Lambda W^TR^TM\omega\\
 &=-y_m^TW^TR^T\alpha^{-1}\omega+\sum_{i=2}^{n}y_{mi}Q_{vi}^TR^T\alpha^{-1}\omega-k_2\sum_{i=2}^{n}{c_{mi}z_{mi}^2}\\
 &~~-\beta_mk_1k_3\sum_{i=2}^n{c_{mi}\lambda_iy_{mi}^2}-k_1k_2\sum_{i=2}^n{ c_{mi}z_{mi}Q_{vi}^TR^TM\omega}\\
 &~~-(1+\beta_m)k_1^2k_3\sum_{i=2}^{n}{\Big(c_{mi}\lambda_iy_{mi}Q_{vi}^TR^TM\omega\Big)}\\
 &=-y_m^TW^TR^T\alpha^{-1}\omega-k_2\sum_{i=2}^{n}{c_{mi}z_{mi}^2}-\beta_mk_1k_3\sum_{i=2}^n{c_{mi}\lambda_iy_m^2}\\
 &~~-k_1k_2\sum_{i=2}^n{\Big( c_{mi}z_{mi}Q_{vi}^TR^TM\omega\Big)}\\
 &~~+\sum_{i=2}^n{y_{mi}\big(Q_{vi}^TR^T\alpha^{-1}-Q_{vi}^T(1+\beta_m)k_1^2k_3c_{mi}\lambda_iR^TM\big)\omega}.
\end{align*}
\end{subequations}
}
By the following inequalities
\begin{subequations}
 \begin{align*}
  &k_1k_2c_{mi}z_{mi}Q_{vi}^TR^TM\omega \leq 
  \frac{\omega^T\epsilon D\omega}{8(n-1)}+r_m z_{mi}^2,
 \end{align*}
\end{subequations}
where $r_z=2(n-1)(k_1k_2c_{mi})^2Q_{vi}^TR^T(D\epsilon)^{-1}M^2RQ_{vi}$,
and 
{\small
\begin{subequations}
\begin{align*}
&y_{mi}Q_{vi}^TR^T\big(\alpha^{-1}-(1+\beta_m)k_1^2k_3c_{mi}\lambda_iM\big)\omega 
\leq \frac{\omega^T\epsilon D\omega}{8(n-1)}+r_z y_{mi}^2,
\end{align*}
\end{subequations}
}
where
{\small
 $$r_y=2(n-1)Q_{vi}^TR^T\big((\alpha^{-1}-(1+\beta_m)k_1^2k_3c_{mi}\lambda_iM\big)^2(\epsilon D)^{-1}RQ_{vi},$$}
we obtain 
{\small
\begin{subequations}
 \begin{align*}
 \dot{V}_m&\leq -y_m^TW^TR^T\alpha^{-1}\omega+\frac{1}{4}\omega^T\epsilon D\omega-\sum_{i=2}^{n}{z_{mi}^2(k_2c_{mi}-r_z)}\\
 &~~-\sum_{i=2}^n{y_{mi}^2(\beta_mk_1k_3c_{mi}\lambda_i-r_y)}.\\
 \end{align*}
\end{subequations}
}
The derivative of $V_d$ is 
 \begin{subequations}
 \begin{align*}
 \dot{V}_d&=-c_dk_2z_d^Tz_d-y_d^T(c_d\beta_dk_1k_3\Lambda)y_d\\
 &~~+z_d^T((1+\beta_d)c_dk_1)W^TR^TD\omega\\
 &~~-y_d^T(\beta_dc_dk_1I_{n-1})W^TR^TD\omega\\
 &=-y_d^TW^TR^T\alpha^{-1}\omega+y_d^TW^TR^T\alpha^{-1}\omega-c_dk_2z_d^Tz_d\\
 &~~-y_d^T(c_d\beta_dk_1k_3\Lambda)y_d+z_d^T((1+\beta_d)c_dk_1)W^TR^TD\omega\\
 &~~-y_d^T(\beta_dc_dk_1)W^TR^TD\omega\\
 &=-y_d^TW^TR^T\alpha^{-1}\omega-c_dk_2z_d^Tz_d-y_d^T(c_d\beta_dk_1k_3\Lambda)y_d\\
 &~~+z_d^T((1+\beta_d)c_dk_1)W^TR^TD\omega\\
 &~~+y_d^TW^TR^T(\alpha^{-1}-\beta_dc_dk_1D)\omega.
 \end{align*}
\end{subequations}
By $\lambda_{\min}\leq \lambda_i$ for all $i=2,\cdots,n$, and inequalities 
\begin{subequations}
 \begin{align*}
  &z_d^TW^TR^T((1+\beta_d)c_dk_1) D\omega\leq \frac{1}{2}\omega^TD\omega^T+\frac{1}{2}z_d^TX_zz_d,
   \end{align*}
\end{subequations}
where $X_z=W^TR^T((1+\beta_d)c_dk_1\alpha D)^2D^{-1} RW$, and 
\begin{subequations}
 \begin{align*}
  &y_d^TW^TR^T(\alpha^{-1}-\beta_dc_dk_1 D)\omega\leq \frac{1}{2}\omega^T(D-\epsilon D)\omega+\frac{1}{2}y_d^TX_yy_d,
 \end{align*}
\end{subequations}
where $X_y=W^TR^T(\alpha^{-1}-\beta_dc_dk_1 D)^2(D-\epsilon D)^{-1}RW$,
we derive,
\begin{subequations}
 \begin{align*}
 \dot{V}_d&\leq-y_d^TW^TR^T\alpha^{-1}\omega-c_dk_2z_d^Tz_d-c_d\beta_dk_1k_3\lambda_{\min}y_d^Ty_d\\
 &~~+z_d^TW^TR^T((1+\beta_d)c_dk_1) D\omega\\
 &~~+y_d^TW^TR^T(\alpha^{-1}-\beta_dc_dk_1 D)\omega\\
 &\leq -y_d^TW^TR^T\omega+\omega^TD\omega-\frac{1}{2}\omega^T\epsilon D\omega\\
 &~~-z_d^T\big(c_dk_2-\frac{1}{2}X_z\big)z_d-y_d^T\big(c_d\beta_dk_1k_3\lambda_{\min}-\frac{1}{2}X_y\big)y_d.
 \end{align*}
\end{subequations}
We consider the following Lyapunov function candidate,  
\begin{subequations}
 \begin{align*}
  V=V_0+V_1+V_m+V_d. 
 \end{align*}
\end{subequations}
In the following, we prove that (i) $\dot{V}\leq 0$, (ii) equilibrium $z^*=(\varphi^*,\omega^*,\rho_1^*,\sigma_1^*,\rho_m^*,\sigma_m^*,\rho_d^*,\sigma_d^*)$ is a strict minimum of $V(\cdot)$ such that
$\nabla V|_{z^*}=0$ and $\nabla^2 V|_{z^*}>0$, and 
(iii) $z^*$ is the only isolated equilibrium in the invariant 
set $\{z\in\Phi\times\mathbb{R}^n\times\mathbb{R}\times\mathbb{R}\times\mathbb{R}^{m-1}\times\mathbb{R}^{m-1}\times\mathbb{R}^{m-1}\times\mathbb{R}^{m-1}|\dot{V}(z)=0\}$ according to Theorem (\ref{thm1}). 

(i). $V$ has derivative
\begin{eqnarray}
 \dot{V}\leq G_0+G_1+G_m+G_d,\label{Lyapunov-derivative}
\end{eqnarray}
where the terms $G_0, G_1, G_m, G_d$ are mainly from $\dot{V}_0, \dot{V}_1, \dot{V}_m, \dot{V}_d$ respectively by leaving the negative terms $-y_m^TW^TR^T\alpha^{-1}\omega$,  $-y_d^TW^TR^T\alpha^{-1}\omega$. $G_0, G_1, G_m, G_d$ have the following forms
\begin{subequations}
 \begin{align*}
 G_0&=-\frac{1}{8}\omega^T\epsilon D\omega,\\
 G_1&=-(k_2\sigma_1-k_2\sigma_1^*)^2-\frac{k_2}{k_1}(k_1\upsilon+k_2\sigma_1)^2,\\
 G_m&=-\sum_{i=2}^{n}{z_{mi}^2(c_{mi}k_2-r_z)}-\sum_{i=2}^n{y_{mi}^2(\beta_mk_1k_3c_{mi}\lambda_i-r_y)},\\
 G_d&=-z_d^T(c_dk_2-\frac{1}{2}X_z)z_d-y_d^T(c_d\beta_dk_1k_3\lambda_{\min}-\frac{1}{2}X_y)y_d.
 \end{align*}
\end{subequations}
It is obvious that $G_0\leq 0$ and $G_1\leq 0$. In the following, we first focus on $G_m$. If there exist $c_{mi}$ and $\beta_m$ such that 
$c_{mi}k_2-r_z>0$ and $\beta_mk_1k_3c_{mi}\lambda_i-r_y>0$, we have $G_m\leq 0$ for all $z_m$ and $y_m$. 
We verify that such $c_{mi}$ and $\beta_m$ do exist. 
By (\ref{eq:R1}) and (\ref{eq2:symmetrizable2}), we have
\begin{subequations}
 \begin{align*}
  Q_{vi}^TR^T\alpha^{-1}RQ_{vi}=Q_{vi}^T\alpha^{-1}_{R}Q_{vi}=1.
 \end{align*}
\end{subequations}
So we only need to prove there exist $c_{mi}$ and $\beta_m$ such that
\begin{subequations}
 \begin{align*}
  Q_{vi}^TR^T(c_{mi}k_2\alpha^{-1})RQ_{vi}-r_z>0,\\
  Q_{vi}^TR^T(\beta_mk_1k_3c_{mi}\lambda_i\alpha^{-1})RQ_{vi}-r_y>0,
 \end{align*}
\end{subequations}
which yields
{\footnotesize
\begin{subequations}\label{inequality1}
 \begin{align}
 c_{mi}k_2\alpha^{-1}&>2(n-1)(c_{mi}k_1k_2M)^2(\epsilon D)^{-1},\\
 (\beta_mk_1k_3c_{mi}\lambda_i\alpha^{-1})&>2(n-1)(\alpha^{-1}-c_{mi}(1+\beta_m)k_1k_3\lambda_iM)^2(\epsilon D)^{-1},
 \end{align}
\end{subequations}
}
From (\ref{inequality1}), we derive
\begin{subequations}
 \begin{align*}
  c_{mi}I_{m-1}&<\frac{\epsilon D}{2(n-1)k_1^2k_2M^2\alpha},\\
  \frac{I_{m-1}}{2a+b+\sqrt{4ab+b^2}}<c_{mi}I_{m-1}&<\frac{2a+b+\sqrt{4ab+b^2}}{2a^2},
 \end{align*}
\end{subequations}
where $a=(1+\beta_m)k_1^2k_3\lambda_iM\alpha),~b=\frac{\beta_mk_1k_3\alpha\lambda_i\epsilon D}{2(n-1)}$.
There exists $c_{mi}>0$ satisfying  (\ref{inequality1}) if 
 \begin{subequations}
  \begin{align*}
    \Big(\frac{I_{n-1}}{2a+b+\sqrt{4ab+b^2}}\Big)_{\max}<\Big(\frac{\epsilon D}{2(n-1)k_1^2k_2M^2\alpha}\Big)_{\min}
  \end{align*}
 \end{subequations}
which can be satisfied by choosing a large $\beta_m$. This is because 
\begin{subequations}
 \begin{align*}
    \lim_{\beta_m\rightarrow \infty}{\Big(\frac{1}{2a+b+\sqrt{4ab+b^2}}\Big)_{\max}}=0, 
 \end{align*}
\end{subequations}
while the term $\Big(\frac{\epsilon D}{2(n-1)k_1^2k_2M^2\alpha}\Big)_{\min}$ does not depend on $\beta_m$. 
Hence there exist $c_{mi}>0$ and $\beta_m>0$ satisfying (\ref{inequality1}) and  $G_m\leq 0$ has been proven. 
Here $(\cdot)_{\max}$ 
and $(\cdot)_{\min}$ are as defined in Assumption (\ref{assumption_coefficient}). 

In the following,  we focus on $G_d$. If there exist $c_d$ and $\beta_d$ such that 
\begin{subequations}
 \begin{align*}
  c_dk_2I_{m-1}-\frac{1}{2}X_z>0,\\
  c_d\beta_dk_1k_3\lambda_{\min}I_{m-1}-\frac{1}{2}X_y>0,
 \end{align*}
\end{subequations}
then $G_d\leq 0$. We prove such $c_d$ and $\beta_d$ do exist with Assumption (\ref{assumption_coefficient}). 
By (\ref{eq:R1}) and (\ref{eq2:symmetrizable2}), we derive 
\begin{subequations}
 \begin{align*}
  W^TR^T\alpha^{-1}RW=W^T\alpha^{-1}_RW=I_{m-1}.
 \end{align*}
\end{subequations}
So we only need to prove there exist $c_d$ and $\beta_d$ such that 
\begin{subequations}
 \begin{align*}
  W^TR^T(c_dk_2\alpha^{-1})RW-\frac{1}{2}X_z>0,\\
  W^TR^T(c_d\beta_dk_1k_3\lambda_{\min})RW-\frac{1}{2}X_y>0,
 \end{align*}
\end{subequations}
which yields
\begin{subequations}\label{inequalities_cd}
 \begin{align}
  c_dk_2\alpha^{-1}&>\frac{1}{2}(1+\beta_d)^2(c_dk_1 D)^2D^{-1},\\
  c_d\beta_d k_1k_3\lambda_{\min}\alpha^{-1}&>\frac{1}{2}(\alpha^{-1}-\beta_dc_dk_1 D)^2(D-\epsilon D)^{-1}.
 \end{align}
\end{subequations}
In the following, we prove that with assumption (\ref{assumption_coefficient}), 
there exist $c_d$ and $\beta_d$ satisfying the above two inequalities (\ref{inequalities_cd}). 
We derive from (\ref{inequalities_cd}) that 
\begin{subequations}
 \begin{align*}
  c_dI_{m-1}< \frac{2k_2}{(1+\beta_d)^2k_1^2\alpha D}~ ,\\
  \frac{1}{{\beta_dk_1\alpha D b_d}}<c_dI_{m-1}<\frac{b_d}{\beta_dk_1\alpha D}~ ,
  \end{align*}
\end{subequations}
where $$b_d=1+k_3\lambda_{\min}(1-\epsilon)+\sqrt{k_3^2\lambda_{\min}^2(1-\epsilon)^2+2k_3\lambda_{\min}(1-\epsilon}).$$
There exists a $c_d$ satisfying the two inequalities (\ref{inequalities_cd}) if there exists a $\beta_d$ such that
\begin{eqnarray}
  \frac{1}{{\beta_dk_1(\alpha D)_{\min} b_d}}<\frac{2k_2}{(1+\beta_d)^2k_1^2(\alpha D)_{\max}}.\label{inequality_betad}
\end{eqnarray}
Since 
\begin{subequations} 
 \begin{align*}
 \lim_{\epsilon\rightarrow 0}{b_d(\epsilon)}&=1+k_3\lambda_{\min}+\sqrt{k_3^2\lambda_{\min}^2+2k_3\lambda_{\min}}\\
  &>1+2k_3\lambda_{\min}, 
 \end{align*}
\end{subequations}
there exists a small $\epsilon>0$ such that 
$b_d(\epsilon)>1+2k_3\lambda_{\min}$. Subsequently (\ref{inequality_betad}) can be 
satisfied if 
\begin{eqnarray}
  \frac{1}{{\beta_dk_1(\alpha D)_{\min}(1+2k_3\lambda_{\min})}}<\frac{2k_2}{(1+\beta_d)^2k_1^2(\alpha D)_{\max}}.\label{inequality_betad2}
\end{eqnarray}
With assumption (\ref{assumption_coefficient}), we can obtain that there exist $\beta_d>0$ satisfying (\ref{inequality_betad2}). Hence
$G_d<0$ is proven and $\dot{V}\leq 0$ subsequently. 

(ii). We prove that the equilibrium $z^*$ is a strict minimum of $V(\cdot)$.
It can be easily verified that $V|_{z^*}=0$ and 
\begin{eqnarray*}
 \nabla V|_{z^*}=\text{col}(\nabla_{\varphi}V,\nabla_{\tilde{z}}V)|_{z^*}=0,
\end{eqnarray*}
where $\tilde{z}=(\omega, \rho_1-\rho_1^*,\sigma_1-\sigma_1^*,\rho_m,\sigma_m,\rho_d,\sigma_d)$. 
Here, we have used $(\omega^*,\rho_m^*,\sigma_m^*,\rho_d^*,\sigma_d^*)=0$. 
The Hessian matrix of $V$ at $z^*$ is 
\begin{eqnarray*}
 \nabla^2V|_{z^*}=\text{blkdiag}(L_p,H),
\end{eqnarray*}
where $\text{bkldiag}$ denotes a block diagonal matrix, $L_p$ is Hessian matrix of $V$ respect to $\varphi$ with $\varphi_1=0$ and 
$H$ is the Hessian matrix of $V$ respect to $\tilde{z}$. It follows \cite[Lemma 5.3]{PIAC}
that $L$ is positive definite. Since the components in $V$ related to $\tilde{z}$ are all 
quadratic and positive definite, $H$ is positive definite. Thus $\nabla^2V|_{z^*}>0$.

(iii). The equilibrium is the only isolated one in the invariant 
set $\{(\varphi,\omega,\eta_m,\xi_m,\eta_d,\xi_d)|\dot{V}=0\}$. Since $\dot{V}=0$,
it yields from (\ref{Lyapunov-derivative}) that $\tilde{z}=0$. Hence 
$\varphi_i$ are all constant. By Proposition (\ref{proposition_equilibrium}),
it follows that $z^*$ is the only isolated equilibrium in the invariant set.

 In this case, $z^*$ is the only one equilibrium in the neighborhood of $z^*$ such that
$\Psi^d=\{(\varphi,\tilde{z})|V(\varphi,\tilde{z})\leq c, \varphi\in\Phi\}$ for some 
$c>0$. Hence with any initial state $z^0$ that satisfies the algebraic equations (\ref{eq:closed_DPIAC2d}),  the 
trajectory converges to the equilibrium state $z^*$.
\hfill $\Box$ 

\begin{remark}\label{GBPIAC_assumption_relaxed}
 In GBPIAC, the dynamics of $(\rho_m,\sigma_m,\rho_d,\sigma_d)$ vanish and the one of 
$(\rho_1,\sigma_1)$ is left only. In the Lyapunov function $V(\cdot)$, with $V_m=0$ and $V_d=0$, it can be proven that the equilibrium of GBPIAC is locally asymptotically stable without Assumption (\ref{assumption_coefficient}). 
\end{remark}

\section{Case study}\label{Sec:case}

\begin{figure}[ht]
 \begin{center}
\includegraphics[scale=0.3]{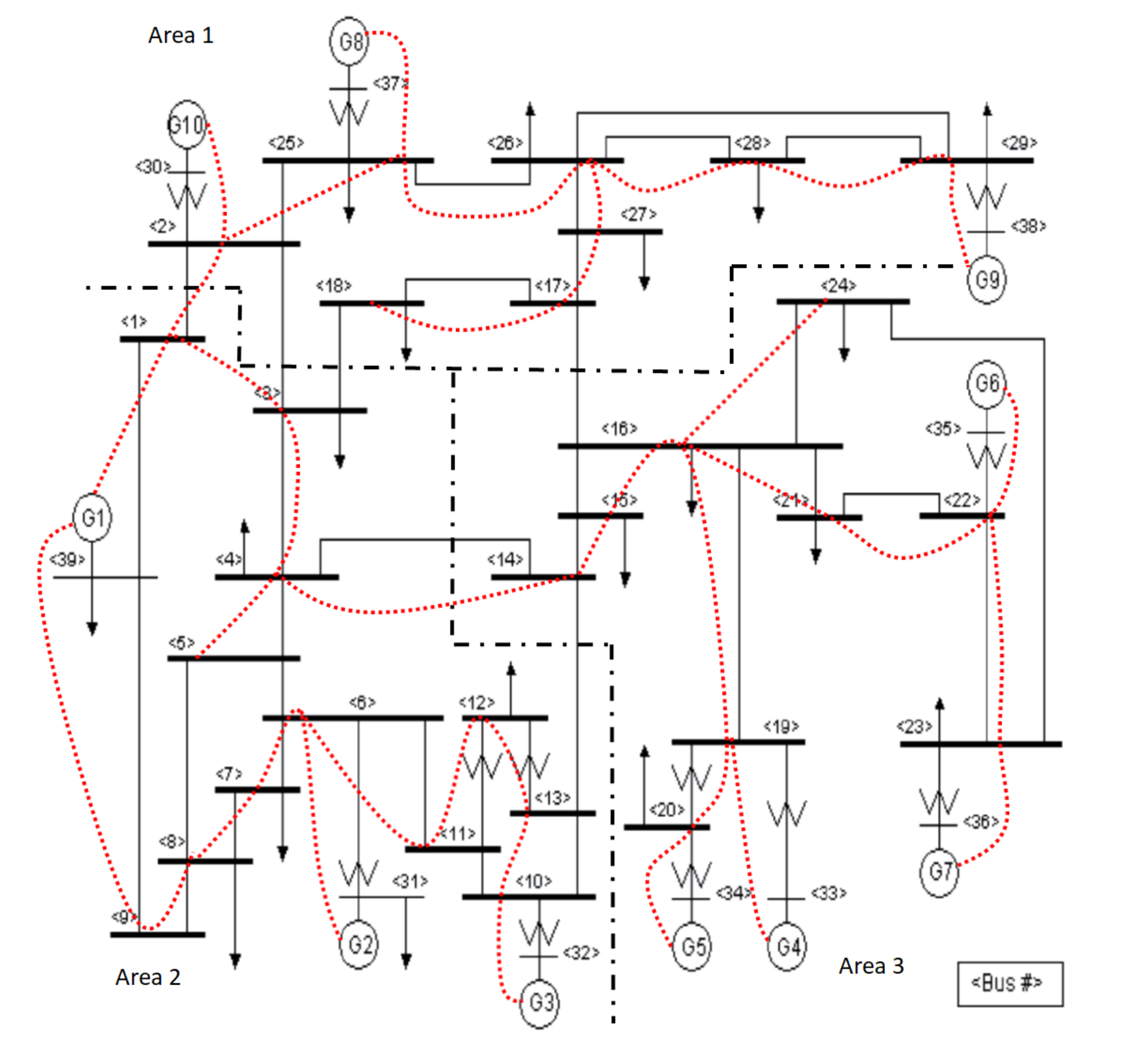}
 \caption{IEEE New England test power system}
  \label{fig.IEEE39_graph}
 \end{center}
\end{figure}

In this section, we evaluate the performance of MLPIAC
on the IEEE-39 buses system as shown in Fig.~\ref{fig.IEEE39_graph}.
We compare MLPIAC to PIAC. For the comparison 
of PIAC to the traditional integral control laws, we refer to \cite{PIAC3}.
In order to study the trade-off between centralized and 
distributed control, we also compare MLPIAC with its two special cases, GBPIAC (\ref{eq:controlPIAC3}) and DPIAC (\ref{eq:DPIAC}).
The performance of the sub-processes identified in Section \ref{Sec:Multi} 
will be observed to study how the transient performance 
is improved in MLPIAC. 
The data of the test system are from \cite{Athay1979}. 
The system consists of 10 generators, 39 buses, which 
serves a total load of about 6 GW. The voltage at each bus is constant
which is derived by power flow calculation with the \emph{Power System Analysis Toolbox} (PSAT) \cite{Milano2008}. There are 49 nodes
in the network including 10 nodes of generators and 39 nodes of buses. In order to 
involve the frequency dependent power sources, we change the buses 
which are not connected to synchronous machines into frequency dependent loads. 
Hence $\mathcal{V}_{M}=\{G1,G2,G3,G4,G5,G6,G7,G8,G9,G10\}$, $\mathcal{V}_{P}=\{30,31,32,33,34,35,36,37,38,39\}$ 
and the other nodes are in set $\mathcal{V}_{F}$. The nodes 
in $\mathcal{V}_M\cup\mathcal{V}_F$ are all equipped with secondary frequency controllers such 
that $\mathcal{V}_K=\mathcal{V}_M\cup\mathcal{V}_F$. 
We remark that each synchronous machine is connected to a bus and its phase angle is rigidly tied 
to the rotor angle of the bus if the voltages of the system are constants \cite{Ilic2000}. Thus the angles of the synchronous machine and the bus 
have the same dynamics. The droop control coefficients are set to $D_i=70$ (p.u./p.u. frequency deviation) 
for all $i\in\mathcal{V}_M\cup\mathcal{V}_F$ under the power base 100 MVA and frequency base 60 Hz. The setting of $D_i$ leads 
to a frequency response of -1.2 p.u./0.1 Hz which equals to that of Quebec power grid connected by bus 39 \cite{balancing_frequency_control}. 
The economic power dispatch coefficient $\alpha_i=1/\beta_i$ where $\beta_i$ is generated randomly with a uniform distribution on $(0,1)$. 
In the simulations, the setting of $\{D_i,~i\in\mathcal{V}_K\}$ and randomly
generated coefficient $\{\alpha_i,~i\in\mathcal{V}_K\}$ yield that 
$(\alpha D)_{\min}=70$ and $(\alpha D)_{\max}=42560$. 
The communication network is assumed to be a spanning tree network as shown by the red lines in Fig.~\ref{fig.IEEE39_graph}, which satisfies 
the requirement in Assumption (\ref{assumption_comm}).
For GBPIAC, the entire network is controlled as a single area. 
For DPIAC, each node $i\in\mathcal{V}_K$ is controlled as a single area.
For MLPIAC, the network is divided into three areas by the dash-dotted black lines as 
shown in Fig.~\ref{fig.IEEE39_graph}.

We set $l_{ij}=1$ if node $i$ and $j$ are connected by a communication line in 
DPIAC (\ref{eq:DPIAC}) and set $l_{rq}=1$ if area $r$ and $q$ are connected by
a communication line in MLPIAC (\ref{multi_level}). 
Note that area 1 and 2 are connected by communication line $(1,2)$ and area 1 and 3 are connected by $(4,14)$. However, area 1 and 3 are not connected by a communication line directly. So their
marginal costs cannot be exchanged. With the control prices $\alpha$ and the Laplacian matrix 
of the communication network, it can be computed 
that $\lambda_{\min}=0.0365$ for DPIAC and $\lambda_{\min}=0.1933$ for MLPIAC.

At the beginning of the simulations, 
the power generation and loads are balanced with nominal frequency $f^*= 60$.  
At time $t=5$ second, the loads at each of the buses 4, 12 and 20 increase 66 MW step-wisely, which 
amounts to a total power imbalance of 198 MW and causes the frequency to 
drop below the nominal 
frequency.


In the following, we evaluate the performance of the control approaches on restoring the nominal 
frequency with a minimized control cost. Because
both GBPIAC and PIAC are centralized control, the principles
of them on improving the transient performance are the same, i.e.,
tuning the corresponding gain coefficients in the decomposed first three sub-processes. Note 
that besides the three sub-processes shared 
with GBPIAC, i.e.,
the convergence processes of $u_s(t)$ to $-P_s$, $\omega_s(t)$ to zero, and the synchronization 
process of $\omega_i$ to $\omega_s(t)$, DPIAC considers the consensus process of the marginal costs of all the nodes and MLPIAC considers the consensus process of the marginal costs of all the areas. We consider the dynamics of the frequency $\overline{\omega}_i(t)=\omega_i(t)+f^*$ and abstract frequency 
$\overline{\omega}_s(t)=\omega_s(t)+f^*$ instead of $\omega_i(t)$ and $\omega_s(t)$ respectively. 
Here, the response
of $\omega_s$ is obtained from (\ref{eq:global}) with $u_s$ as the total 
control input of the three methods respectively. For illustrations of the extra frequency  oscillation caused by the overshoot of $u_s$, we refer to the simulations in, e.g.,
\cite{Dorfler2014,EAGC,Trip2016240,PIAC3,Zhao2015,PIAC3}.

\begin{figure*}[ht]
\begin{center}
\includegraphics[scale=1.16]{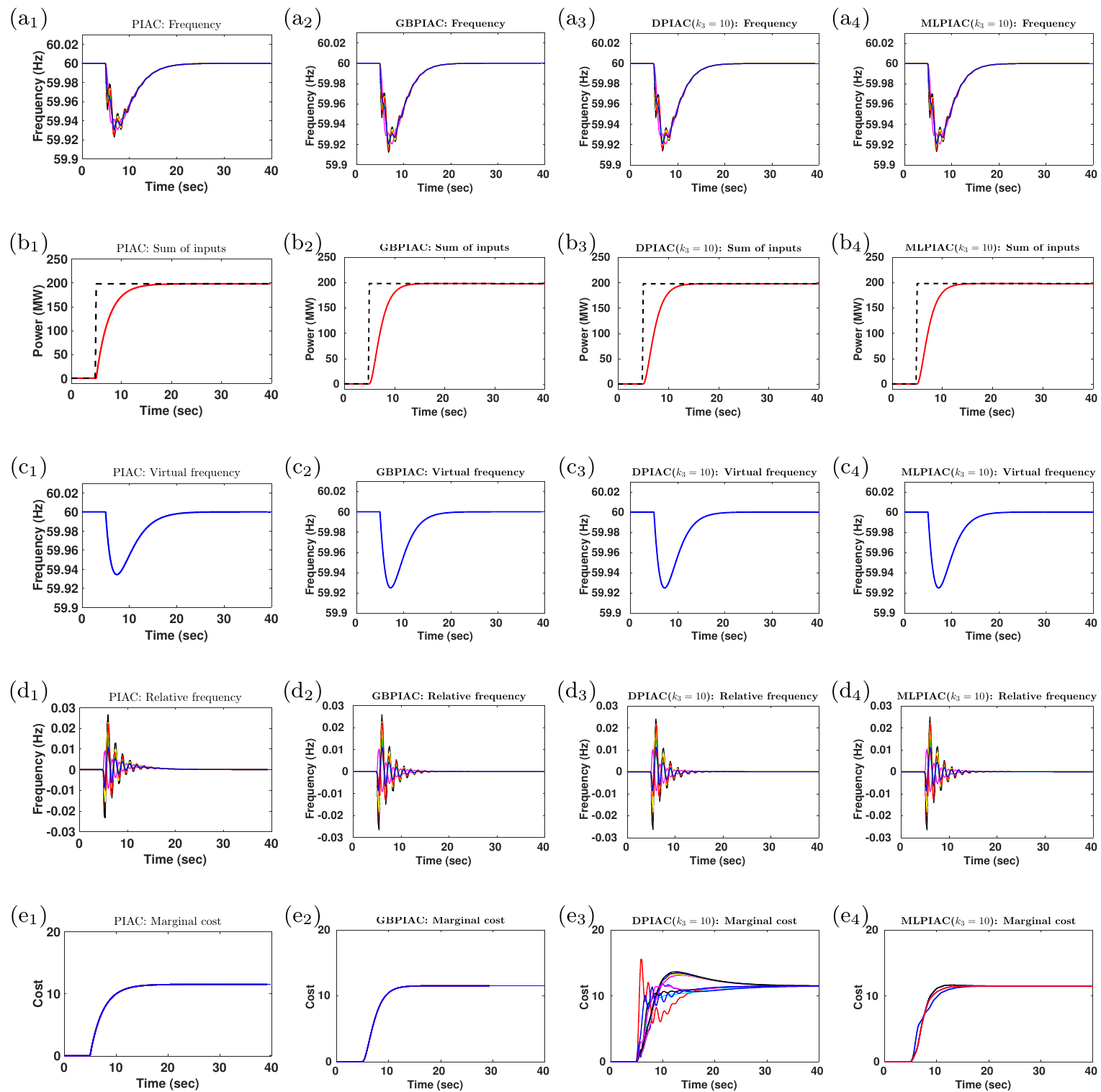}
\caption{The simulation results of the PIAC, GBPIAC, DPIAC and MLPIAC methods on IEEE 39-bus test system. The control coefficient is set $k=0.4$ in PIAC, and $k_1=0.4, k_2=1.6$ in
the other three methods.
The black dashed lines in b$_1$-b$_4$, denote the power imbalance of the network.}
\label{fig:basic}
\end{center}
\end{figure*}
%

The simulation results are shown in Fig.~\ref{fig:basic} where there are 20 plots in 5 rows and 4 columns. 
The plots in the rows from top to bottom illustrate the dynamics of the frequency $\overline{\omega}_i(t)\in\mathcal{V}_M$,
control input $u_s(t)$, abstract frequency $\overline{\omega}_s(t)$, 
relative frequency $\omega_i(t)-\omega_s(t)$ for all $i\in\mathcal{V}_M$, and
 marginal costs of the controllers at the nodes of the synchronous machines in DPIAC and of the areas in MLPIAC, and the plots
in the column from left to right illustrate the results of PIAC, GBPIAC, DPIAC with $k_3=10$, and MLPIAC  with $k_3=10$ respectively. In these four simulations, we set $k=0.4$ for PIAC, $k_1=0.4,k_2=1.6$ for GBPIAC, DPIAC and MLPIAC. Note
that Assumption (\ref{assumption_coefficient}) is not satisfied in the simulations of DPIAC and MLPIAC, i.e., $\frac{k_2}{k_1}=4$
while the values of $\frac{2(\alpha D)_{\max}}{(\alpha D)_{\min}+2k_3\lambda_{\min}}$ are about 1203 and 1152 for DPIAC and MLPIAC with $k_3=10$ respectively.
We remark that the relative frequency describes the 
synchronization process of $\omega_i(t)$ to $\omega_s(t)$, which 
is the main concern of primary control.

Let us first focus on the performances of PIAC and GBPIAC in the plots 
in the first and second column. It can be observed from the plots from top to bottom that the frequencies are restored to the nominal frequency
without any extra oscillations, the control input converges to the power imbalance 
exponentially without an overshoot, the abstract frequency converges to the 
nominal frequency without an overshoot, the frequencies synchronize to 
the abstract frequency $\omega_s$, the marginal costs 
are identical at the nodes of the synchronous machines. It can 
be easily observed that the performance of GBPIAC
is similar to that of PIAC method, which indicates 
that the transient performance of the frequency can  be improved by GBPIAC.

Second, we turn to the performance of DPIAC in the plots in the third 
column. Compared with GBPIAC by observing the plots from top to bottom,
the dynamics of $\overline{\omega}_i$ are similar to that of GBPIAC, the control input $u_s(t)$ and $\overline{\omega}_s(t)$ are identical to that in GBPIAC, the synchronization 
of $\omega_i$ to $\omega_s$ is similar to GBPIAC with slightly smaller magnitude 
of oscillations, the marginal
 costs achieve a consensus at the steady state. However,
 the marginal costs are not identical in the transient phase, which is different from that in GBPIAC.

Third, we consider the performance of MLPIAC. It can be seen from Fig.~\ref{fig:basic}a$_4$, Fig.~\ref{fig:basic}b$_4$, 
Fig.~\ref{fig:basic}c$_4$ and Fig.~\ref{fig:basic}d$_4$ that, the subprocesses of $u_s(t)$, $\omega_s(t)$ and $\omega_i-\omega_s$ 
are similar as in GBPIAC and DPIAC. However, the marginal costs of the three areas shown in Fig.~\ref{fig:basic}e$_4$ achieve consensus 
much faster than in DPIAC even though the control gain $k_3$ for the consensus process equals to the one in DPIAC. 
Hence, for a large-scale network, \emph{the multi-level control is more effective in decreasing the control cost than the distributed method.} So MLPIAC 
balances the advantages and disadvantages of the centralized and distributed control.  

\section{Conclusion}\label{Sec:conclustion}

In this paper, we proposed a multi-level secondary frequency control, called 
\emph{Multi-Level Power-Imbalance Allocation control} (MLPIAC), for a large-scale power system
with multiple areas. Centralized 
control is implemented within each area and distributed
control is implemented over the areas. At the steady state,
the nominal frequency is restored with a minimized control cost. At the 
transient phase, the system performance can be improved 
by tuning the control gain coefficients without extensive computations.
A trade-off between centralized control and distributed control is determined. The
 minimal cost can be more effectively achieved
than the purely distributed control due to 
the smaller number of areas than that of nodes in the network.
The requirements on communications and computations are reduced
compared to pure centralized control due to the smaller number of nodes in 
each area than that of nodes in the network. 

However, Assumption (\ref{assumption_coefficient})
is still required for the asymptotic stability of MLPIAC even 
though it is not required in the numerical simulations. How to relax Assumption (\ref{assumption_coefficient})
theoretically still needs further consideration. The control law may 
be applicable for the power systems with general convex cost functions, which 
however needs further study specially in the asymptotic stability analysis. 
Furthermore, there usually are time-delays and noises in the measurement 
of the frequency and communications in practice, in which case the robustness 
of MLPIAC needs to be evaluated.

\section*{Acknowledgment}

The authors thank Dr. Johan L. A. Dubbeldam from Delft University of Technology,
Prof. Tongchao Lu from Shandong University, 
Mr. C\'{e}sar A. Uribe from University of Illinois
for their useful discussions
on this research, including the control algorithm synthesis and 
stability analysis. Kaihua Xi thanks the China Scholarship Council for the financial support.

\appendices


\section{Notations}\label{appendix_notation}


Denote the number of nodes in the sets $\mathcal{V}_M,\mathcal{V}_F,\mathcal{V}_P$ and $\mathcal{V}_K$ by
$n_m,n_f,n_p,n$ respectively, the total number of nodes in the power system by $n_t$.
So $n=n_m+n_f$ and $n_t=n_m+n_f+n_p$. 

To express simply, we write a diagonal matrix $\beta=\text{diag}{(\{\beta_i,i\cdots n\})}\in\mathbb{R}^{n\times n}$ with $\beta_i\in\mathbb{R}$ as 
$\text{diag}{(\beta_i)}$. 
It is convenient to introduce the matrices
$D=\text{diag}(D_i)\in\mathbb{R}^{n\times n}$, 
$M=\text{diag}(M_i)\in\mathbb{R}^{n\times n}$, $\alpha=\text{diag}(\alpha_i)\in\mathbb{R}^{n\times n}$.
Note that $M_i=0$ for $i\in\mathcal{V}_F$ and $M_i>0$ for $i\in\mathcal{V}_M$.  
Denote the identity matrix by $I_n\in\mathbb{R}^{n\times n}$, the $n$ dimension vector with all elements equal to one by $1_n$, 
$\theta=\text{col}(\theta_i)\in\mathbb{R}^{n_t}$, $\omega=\text{col}(\omega_i)\in\mathbb{R}^n$, $P=\text{col}(P_i)\in\mathbb{R}^{n_t}$
and $\varphi=\text{col}(\varphi_i)\in\mathbb{R}^{n_t}$ where $\varphi_i=\theta_i-\theta_1$ for all $i\in\mathcal{V}$, $\eta=\text{col}(\eta_i)\in\mathbb{R}^n$, and $\xi=\text{col}(\xi_i)\in\mathbb{R}^n$.

The number of the areas is denoted by $m$, the control price of area $r$ is $\alpha_r$. Denote 
$\alpha_R=\text{diag}(\alpha_r)\in\mathbb{R}^{m\times m}$. Define matrix $R=(r_{ir})\in\mathbb{R}^{n\times m}$
with $r_{ir}=1$ if node $i$ belongs to area $r$, otherwise $r_{ir}=0$. Note that $\alpha,\alpha_{R}, R$ satisfy 
\begin{subequations}\label{eq:R}
 \begin{align}
R^T\alpha^{-1}R=\alpha_R^{-1},\label{eq:R1}\\
1_n=R1_m.\label{eq:R2}
 \end{align}
\end{subequations}
Denote $L\in \mathbb{R}^{m\times m}$ as the Laplacian matrix of the communication network as defined in (\ref{eq:laplacian2}). 

For symmetric matrices $A$ and $B$, we say $A>0$ (or $A\geq 0$ if $A$ is positive-definite (or semi-positive-definite), 
and say $A>B$ (or $A\geq B$) if $(A-B)$ is positive-definite (or semi-positive-definite). 

The following 
inequality is used frequently in the following stability analysis of the control laws. For any 
$x\in\mathbb{R}^m,y\in\mathbb{R}^m$, the following inequality holds
\begin{eqnarray*}
 x^T y \leq \frac{1}{2}x^T\varepsilon x+\frac{1}{2}y^T\varepsilon^{-1} y,
\end{eqnarray*}
where $\varepsilon\in\mathbb{R}^{m\times m}$ is an invertible positive-definite diagonal 
matrix. The inequality follows from 
\[\begin{bmatrix}
      x \\
      y
    \end{bmatrix}^T
    \begin{bmatrix}
      \varepsilon&-I_m \\
      -I_m&\varepsilon^{-1}
    \end{bmatrix}
    \begin{bmatrix}
      x \\
      y
    \end{bmatrix}\geq 0.
\]

\section{Preliminaries on DAEs}\label{appendix_DAE}

Consider the following DAE systems
\begin{subequations}\label{eq:DAEtt}
\begin{align}
 \dot{x}&=f(x,y),\label{eq:DAE}\\
  0&=g(x,y), \label{Eq:DAE2}
\end{align}
\end{subequations}
where $x\in \mathbb{R}^n, y\in \mathbb{R}^m$ and $f:\mathbb{R}^n\times \mathbb{R}^m\rightarrow \mathbb{R}^n$ and 
$g:\mathbb{R}^n\times \mathbb{R}^m\rightarrow \mathbb{R}^m$ are 
twice continuously differentiable functions. $(x(x_0,y_0,t),y(x_0,y_0,t))$ is the solution with 
the admissible initial conditions $(x_0,y_0)$ satisfying the algebraic constraints
\begin{align}
 0=g(x_0,y_0), \label{eq:initial}
\end{align}
and the maximal domain of a solution of (\ref{eq:DAEtt}) is denoted by $\mathcal{I}\subset \mathbb{R}_{\geq 0}$ where 
$\mathbb{R}_{\geq 0}=\{x\in \mathbb{R}| x\geq 0\}$. 

Before presenting the Lyapunov/LaSalle stability criterion of the DAE system, we make the following two assumptions.

\begin{assumption}\label{assumption_appendix1}
  The DAE system possesses an equilibrium state $(x^*,y^*)$ such 
that $f(x^*,y^*)=0,~ g(x^*,y^*)=0$. 
\end{assumption}

\begin{assumption}\label{assumption_appendix2}
  Let $\Omega\subseteq \mathbb{R}^n\times \mathbb{R}^m$ be an open connected set containing $(x^*,y^*)$,
assume ({\ref{Eq:DAE2}}) is \emph{regular} such that 
the Jacobian of $g$ with respect to $y$ is a full rank matrix for any $(x,y)\in \Omega$, i.e.,
\begin{align*}
\text{rank}(\nabla_{y}g(x,y))=m, ~\forall (x,y)\in \Omega.
\end{align*}
\end{assumption}

Assumption (\ref{assumption_appendix2}) ensures the existence and uniqueness of the solutions 
of (\ref{eq:DAEtt}) in $\Omega$ over the interval $\mathcal{I}$ with the 
initial condition $(x_0,y_0)$ satisfying (\ref{eq:initial}). 

The following theorem provides a sufficient stability condition for the stability of the equilibrium of DAE in (\ref{eq:DAEtt}).

\begin{theorem}\label{thm1}
(Lyapunov/LaSalle stability criterion \cite{Schiffer,Hill1990}): Consider the DAE system in (\ref{eq:DAEtt}) with 
assumptions (\ref{assumption_appendix1}) and (\ref{assumption_appendix2}), and an equilibrium $(x^*,y^*)\in \Omega_H\subset \Omega$. If there exists 
a continuously differentiable function $H:\Omega_H\rightarrow R$, such 
that $(x^*,y^*)$ is a strict minimum of $H$ i.e., $\nabla H|_{(x^*,y^*)}=0$ and $\nabla^2H|_{(x^*,y^*)}>0$, 
and $\dot{H}(x,y)\leq 0,~\forall (x,y)\in \Omega_H$, then the following 
statements hold:

(1). $(x^*,y^*)$ is a stable equilibrium with a local Lyapunov function 
$V(x,y)=H(x,y)-H(x^*,y^*)\geq 0$ for $(x,y)$ near $(x^*,y^*)$, 

(2). Let $\Omega_c=\{(x,y)\in \Omega_H|H(x,y)\leq c\}$ be a compact sub-level set for 
a $c>H(x^*,y^*)$. If no solution can stay in $\{(x,y)\in \Omega_c|\dot{H}(x,y)= 0\}$ 
other than $(x^*,y^*)$, then $(x^*,y^*)$ is asymptotically stable. 
\end{theorem}
We refer to \cite{Schiffer} and \cite{Hill1990} for the proof of Theorem \ref{thm1}. 

\section{Preliminaries of Symmetrizable matrix}\label{appendixSymmetrizable}

\begin{definition}
 A matrix $B\in\mathbb{R}^{m\times m}$ is \emph{symmetrizable} if 
 there exists a positive-definite invertible diagonal matrix $A\in\mathbb{R}^{m\times m}$ and a symmetric matrix $L\in\mathbb{R}^{m\times m}$ such that $B=LA$. 
\end{definition}
\begin{theorem}\label{theorem:Symmetrizable}
Consider the Laplaciam matrix $L\in\mathbb{R}^{m\times m}$ as defined in (\ref{eq:laplacian2}) and the positive-definite 
diagonal matrix $\alpha_R\in\mathbb{R}^{m\times m}$ as defined in Appendix (\ref{appendix_notation}). The matrix $L\alpha_R$ is 
a symmetrizable matrix and there exists an invertible matrix $Q$ such that 
\begin{eqnarray}\label{diagonalizematrix}
 Q^{-1}L\alpha_R Q=\Lambda,
\end{eqnarray}
where $\Lambda=\text{diag}(\lambda_i)\in\mathbb{R}^{m \times m}$, $\lambda_i$ is the eigenvalue of $L\alpha$ and $\lambda_1=0$.
 Denote $Q^{-1}=[Q_{v1},Q_{v2},\cdots,Q_{vm}]^T$ and $Q=[Q_{1},Q_{2},\cdots,Q_{m}]$, we have 
 \begin{subequations}
  \begin{align}
    Q_{v1}&=1_m, \label{eq:symmetraizable1}\\
    ~\alpha_R Q_1&= 1_m, \label{eq:symmetraizable2}\\
  ~Q^T\alpha_R Q&=I_m,\label{eq:symmetrizable3}\\
  ~Q^{-1}\alpha_R^{-1}(Q^{-1})^T&=I_m, \label{eq:symmetrizable4}\\
  Q^{-1}&=Q^T\alpha_R.\label{eq:symmetrizable5}
  \end{align}
 \end{subequations}

 Furthermore, the new matrix $W=[Q_{v2},\cdots,Q_{vm}]\in\mathbb{R}^{n\times(m-1)}$ and $S=[Q_{2},\cdots,Q_{m}]\in\mathbb{R}^{m\times(m-1)}$
 satisfy that 
 \begin{subequations}
  \begin{align}
    W^TS&=I_{(m-1)}, \label{eq2:symmetrizable1}\\
 W^T\alpha_R^{-1} W&=I_{(m-1)}, \label{eq2:symmetrizable2}\\
 S^T\alpha_R S&=I_{(m-1)}, \label{eq2:symmetrizable3}\\
 Q_{vi}&=\alpha_R Q_i,\\
 W&=\alpha_R S.
  \end{align}
 \end{subequations}
\end{theorem}
\emph{Proof}:
Let $T=\sqrt{\alpha_R}$ which is a diagonal matrix, then 
$$TL\alpha_R T^{-1}=\sqrt{\alpha_R}L\sqrt{\alpha_R}.$$
Hence, there exists an invertible matrix $\Gamma^{-1}=\Gamma^T$ such that 
$$\Gamma^{-1}TL\alpha_R T^{-1} \Gamma=\Gamma^{-1}\sqrt{\alpha_R}L\sqrt{\alpha_R}\Gamma=\Lambda.$$ 
Let $Q=T^{-1}\Gamma$, we derive $Q^{-1}L\alpha_R Q=\Lambda$. Since $L$ is a Laplacian matrix as 
defined in (\ref{eq:laplacian2}),
we have $1_m^TL\alpha_R=0$, then there is a zero eigenvalue, i.e., $\lambda_1=0$.
Denote $\Gamma=[\Gamma_1,\Gamma_2,\cdots,\Gamma_m]$, then $\Gamma^{-1}=\Gamma^{T}=[\Gamma_1,\Gamma_2,\cdots,\Gamma_m]^T$. 

Since $\Gamma_1$ is the eigenvector corresponding to $\lambda_1=0$ of $\sqrt{\alpha_R} L\sqrt{\alpha_R}$ such 
that $L\sqrt{\alpha_R} \Gamma_1=0$, from which 
we derive $\sqrt{\alpha_R}\Gamma_1=1_m$. Hence by $Q=T^{-1}\Gamma$, we obtain $Q_1=(\sqrt{\alpha_R})^{-1}\Gamma_1=\alpha_R^{-1}1_m$. Similarly
we derive $Q_{v1}=\Gamma_1^TT=\Gamma^T\sqrt{\alpha_R}=\sqrt{\alpha_R}\Gamma_1=1_m$.

By $Q=T^{-1}\Gamma$, we derive $\Gamma^T\alpha_R Q=\Gamma^TT^{-1}\alpha_R T^{-1} \Gamma=I_m$. $Q^{-1}\alpha_R^{-1}(Q^{-1})^T=R^TR=I_m$ can be 
obtained similarly.

(\ref{eq2:symmetrizable1}) is yielded directly from $Q^{-1}Q=I_m$. (\ref{eq:symmetrizable3}) and (\ref{eq:symmetrizable4}) yields (
\ref{eq2:symmetrizable2}) and (\ref{eq2:symmetrizable3}). \hfill $\Box$

\begin{small}
\bibliography{ifacconf}         
\bibliographystyle{plain}
\end{small}

\end{document}